\documentclass{jpsj2}

\def\runtitle{Instructions for the Preparation of Manuscript}
\def\runauthor{Online-Journal Subcommittee}

\title{%
Charge Ordered Insulator without Magnetic Order Studied by Correlator Projection Method}

\author{%
Kota {\sc Hanasaki}$^{1}$~\thanks{E-mail address: khanasak@issp.u-tokyo.ac.jp} and Masatoshi {\sc Imada}$^{1, 2}$
}

\inst{%
$^{1}$Institute for Solid State Physics, University of Tokyo, Kashiwanoha 5-1-5, Kashiwa, Chiba 277-8581\\
$^{2}$PRESTO, Japan Science and Technology Agency
}

\recdate{\today}
\abst{%
The Hubbard model with additional intersite interaction `$V$' ( the extended Hubbard model) is investigated by the correlator projection method (CPM). CPM is a newly developed numerical method which combines the equation-of-motion approach and the dynamical mean-field theory.
  Using this method, properties of the extended Hubbard Model at quarter filling are discussed with special emphasis on the metal-insulator transition induced by electron-electron correlations. 
  The result shows a metal-insulator transition to a charge ordered insulator with antiferromagnetic order at low temperatures, but a charge ordered insulator without magnetic symmetry breaking at intermediate temperatures. Here, the magnetic order is found to be suppressed to low temperatures because of the smallness of the exchange coupling $J_{\rm eff}$. The present results are in sharp contrast with the Hatree-Fock approximation whereas in agreement with the experimental results on quarter-filled materials with strong correlations, such as organic conductors. 
}

\kword{%
extended Hubbard model, metal-insulator transition, intersite Coulomb interaction
}

\begin{document}
\maketitle
\section{Introduction}
Translational symmetry breaking in the charge degrees of freedom, or charge ordering, is widely observed in strongly correlated electron systems such as manganites,~\cite{MnCO,MnCO2,MnTokura} vanadates,~\cite{Ueda} and various organic conductors.~\cite{Mori98,DCNQI}
In these systems, the charge-ordering transition often accompany a large jump in electric resistivity or metal-insulator transition (MIT).~\cite{Imada98}
It has attracted much interest both from theoretical and experimental point of views.~\cite{Imada98,Solov,Fujimori,Seo,HubbardEHM}  
Charge ordering is indeed one of the fundamental routes to cause metal-insulator transitions in {\it non}-integer filling systems. The nature of this transition, however, is not well understood yet. 
Here, we refer to {\it non}-integer filling as the state where the {\it electron number} per site is not an integer. 
The insulating phase of these systems usually have translational symmetry breaking in the charge degrees of freedom. 
This is in marked contrast to the case of integer-filling systems, which have integer numbers of electron(s) per site. Insulating phase in the latter systems, such as the Mott insulator phase, usually has no symmetry breaking in the charge degrees of freedom.
Here in this paper, we focus on systems at simple rational fillings that stabilize {\it commensurate} charge orders. We also focus on the interplay of the on-site and intersite Coulomb interactions in such charge-ordering transitions. 

There are several properties in charge-ordering transitions that require theoretical clarifications:\\
1) The system can become insulating with charge order but without the magnetic order: This is expected if one considers the strong-coupling limit. 
Systems with repulsive interaction at commensurate filling in this limit takes the lowest energy when the charge forms a certain spatial pattern to reduce the Coulomb interaction energy. In such phases, any excitations in the charge degrees of freedom has a finite gap. The system is then an insulator irrespective of the magnetic ordering, which may not be formed due to the quantum and/or thermal fluctuations. 
This expectation remains to be clarified except that limit. 
Experimentally, non-magnetic insulating phase is indeed observed at intermediate temperatures in wide range of real materials~\cite{Mori98,DCNQI}.\\
 2) Many systems undergo metal-insulator transitions simultaneously at the point where the long-ranged charge order is formed {\it i.e.} no intermediate phase with charge order and metallic conductivity appears: 
 Here, we do not in principle exclude the possibility of the charge-ordered metallic state. However, the above-mentioned feature is experimentally observed in most of the real charge-ordering materials~\cite{MnTokura,MnKawano, Ueda,UedaCO, Mori98,Miyagawa00}.
 
 On the other hand, using simple theoretical methods such as the mean-field approximation, one easily obtain charge-ordered metallic phase in a wide region of the parameter space as we discuss below. Mean-field approximations fail in reproducing the general tendency in realistic parameter region. 
 
 Studies on metal-insulator transition induced by the strong electron-electron correlations require careful treatment of the two intriguing aspects of quantum mechanics, itineracy and locality. This is one of the most fundamental problems in condensed matter physics. 
 In integer-filling systems, such transition is called the Mott transition~\cite{Mott} and has been the subject of intensive research.~\cite{Imada98}
  These studies were based on a lattice model with only the on-site electron repulsion, called the Hubbard model. 
 The Hubbard model is defined as
\begin{equation}
{\cal H}= -\sum_{i,j,\sigma}t_{i,j} c^{\dagger}_{i\sigma}c_{j\sigma}
+ U\sum_{i}n_{i\uparrow}n_{i\downarrow}.
\label{eqn:HM}
\end{equation}
 Here, $c_{i\sigma}(c^{\dagger}_{i\sigma})$ is the annihilation (creation) operator of an electron at an atomic site $i$ with a spin index $\sigma$. 
 The number operator on site $i$ with spin $\sigma$ is represented by $n_{i\sigma}$, while $t_{i,j}$ is an electron transfer from an atomic site $i$ to $j$, and $U$ is the local Coulomb repulsion.
 Recent development of numerical calculations has allowed detailed analyses on the Mott transition.~\cite{FurukawaQMC,Kashima,Watanabe} The dynamical mean-field theory~\cite{Georges} and its extension have also applied to clarify finite-temperature properties of the Mott transition.~\cite{Zhang,Onoda2} 
  
  On the other hand, as for metal-insulator transitions in non-integer filling systems, theoretical mehods are still not well established. In this paper, we will study on the methods to treat those transitions induced by the charge ordering. 
  
  Charge ordering in a broad sense includes the Wigner crystallization.~\cite{Wigner} The charge ordering in strongly correlated systems, however, is much different from the Wigner crystallization of the electron gas in the continuum space: First, there is a periodic potential formed by the positively ~(or negatively, if carriers are holes)~charged ions, which may allow the ordering at relatively higher carrier concentrations. Second, as a consequence of the first, the ordering occurs mostly at commensurate fillings and easily melts away from it. The importance of commensurability has been pointed out in ref.~\cite{Noda01} In this paper, we call only this commensurate case as the charge order.
 In order to obtain a model for this charge order, we assume that the long-ranged part of the Coulomb potential is screened by the large carrier concentration. We define the simplified model, the extended Hubbard model as
\begin{equation}
  {\cal H}= - \sum_{i,j} t_{i,j} c^{\dagger}_{i \sigma} c_{j \sigma}
  +\sum_{i} U n_{i \uparrow} n_{i \downarrow} 
  + \sum_{(i,j)} V_{i.j} n_{i} n_{j}.
\label{eqn:EHM}
\end{equation}
  Here, $V_{i,j}$ is the off-site Coulomb interaction. Other notation is the same as eq.(\ref{eqn:HM}). For simplicity, we restrict the range of $V$ to only the nearest neighbor sites. 
We consider square lattice and hopping integral $-t_{i,j}$ takes the value $-t$ for only the nearest neighbor pair $(i,j)$ and zero otherwise.

 Charge ordering has been studied by several theoretical methods using the extended Hubbard model defined in eq.~(\ref{eqn:EHM}). 
 Here, we will examine some of the existing theoretical methods.
 The Hartree-Fock approximation (HFA) is the most standard method to treat the charge ordering problems.~\cite{Seo} Using HFA, one can reproduce charge-ordered phase. On the other hand, HFA cannot reproduce the charge-ordered insulating phase without magnetic ordering. HFA also overestimates the ordered phases. The drawbacks arise from the neglect of dynamical and spatial correlation effects.
 Perturbative approaches~\cite{Aoki} may be powerful tool for analyzing metallic phase of these systems. However, most of them cannot describe insulating or symmetry-broken phases.
 The dynamical mean-field theory (DMFT) can also be applied with appropriate mapping to the infinite dimensional model. 
 It, however, fails in reproducing the insulating phase in the absence of the magnetic order.~\cite{Pietig} This is ascribed to the fact that it treats the intersite interactions only in the static mean-field level. 
 Exact diagonalization (ED) can also be a standard approach to this problem.~\cite{Mck01} There is, however, finite size effect that becomes severe in determining metal-insulator transitions and/or charge ordering transitions. It is also difficult to obtain finite-temperature properties.
 
 Above considerations impose the requirements for an improved theoretical approach; (1) a non-perturbative method to describe the insulating and/or charge-ordered phases, (2) an analytic method that can describe the finite-temperature phases, (3) a method which can equally treat both intersite and onsite interactions to realize insulating phases.
 
 In this paper, we analyze the charge-ordering phenomena by using the correlator projection method (CPM).~\cite{Onoda2} We show that it succeeds in reproducing the presence of the insulating charge-ordered phase in the absence of the magnetic order.
 
 The formulation of the correlator projection method is given in Sec.2. Here, we summarize the properties of this method. 
 CPM is a newly developed method that combines the equation-of-motion approach~\cite{Nakajima,Mori,Zwanzig} and the dynamical mean-field theory~\cite{Georges}.
 It can treat the correlation effects beyond the mean-field level in a self-consistent manner. It is also a non-perturbative method that can describe the symmetry broken state and/or the insulating state. 
 It consists of two parts: First, one derives the analytic form of the Green's function by the equation-of-motion approach. Second, one calculates the self-energy part by extending the dynamical mean-field theory in order to restore the fluctuation effects.
 These two parts are performed in a self-consistent manner on the basis of the operator projection theory and the dynamical mean-field theory.  
 
 The dynamical mean-field theory is suited for treating the effects of dynamical quantum fluctuations. However, it may not be adequate to treat that of spatial fluctuations. 
 CPM allows to overcome this drawback by taking account of spatial fluctuations.We concentrate on the system at commensurate filling in strong coupling regime, where the effect of spatial fluctuation is relatively small. 
 
 The results obtained by the present approach are given in Sec.3. CPM have reproduced charge-ordered insulating phase without order in the spin-degree of freedom. 

The result is in agreement with several experimental observations. Comparisons to real materials are given in Sec.4.
 
 Sec.5 is devoted to the summary of this paper. 

\section{Formulation}
\subsection{Correlator Projection Method}
As we have mentioned in the introduction, the correlator projection method~(CPM) combines two procedures.

The first procedure is based on the equation-of-motion decoupling approach originally named the operator projection method~(OPM).~\cite{Nakajima,Zwanzig,Mori} This approach, which was originally developed for analyses of stochastic phenomena, was first applied to the Hubbard model by Roth~\cite{Roth}. 
She solved the equation of motion up to the second order and obtained a Green's function with two poles. This approximation is called the two-pole approximation (TPA). TPA has been further applied for the Hubbard model and some other models with strong electron-electron correlations.~\cite{Edward,Mancini0}

 Equation-of-motion approaches are suited for strongly correlated electron systems since one can treat the local constraint (the Pauli principle) correctly and can distinguish between the double occupancy and single occupancy. Thus one can treat the strong on-site and off-site Coulomb interaction. The obtained Green's function correctly reproduces the high-energy part of physics.   
On the other hand, however, they tend to fail in describing the low-energy part of physics since they decouple the higher-order operators, which describe low-energy collective motions. 

The correlator projection method overcomes this drawback by introducing a refined self-energy at the point one truncates the equation of motion.
When one solves the equation of motion for the Green's function, it leaves the self-energy as an unknown quantity. The equation of motion for the self-energy (, which we call the first-order self-energy $\Sigma^{(1)}$ from now on) is then given and leaves the second-order self-energy $\Sigma^{(2)}$ as an unknown object. This sequence of continued fraction expansion continues. 
Using OPM, one can obtain the explicit form of the self-energy, which we call `the higher-order self-energy', $\Sigma^{(n)}$. The second procedure of CPM is to evaluate the highest-order self-energy, say $\Sigma^{(n)}$, in self-consistent manner by extending the dynamical mean-field theory. 
The continued fraction is truncated at $\Sigma^{(n)}$. Then, $\Sigma^{(n)}$ is self-consistently determined by assuming that $\Sigma^{(n)}$ is momentum independent.~\cite{Onoda2}
This reduces to the original dynamical mean-field theory when one takes $n=1$.
By solving a higher-order self-energy self-consistently, it systematically incorporates the spatial fluctuations with increasing $n$. For practical use, even \(n=2\) gives a significant improvement, where $\Sigma^{(1)}$ recovers momentum dependence.~\cite{Onoda2}

The first attempt of applying this method was given in the half-filled Hubbard model with the second-neighbor hopping $t^{\prime}$~\cite{Onoda2}. It reproduced the metal-insulator transition at the parameters in good agreement with a reliable numerical calculation~\cite{Kashima}. 
 
\subsection{Formulation of CPM}
Now, we summarize the concrete formulation of CPM.~\cite{Onoda2}

\subsubsection{The first procedure; OPM}
One starts with a set of operators $\{ \phi_{i} \}$ for physical quantities.
The equation of motion for the operators is expressed as
\begin{equation}
-\frac{\partial \phi_{i}}{\partial \tau} = \hat{\omega} \phi_{i}.
\label{eqn:OPM1}
\end{equation}
Through the remaining part of this paper, we take imaginary time representation;  \( \phi(\tau) \equiv e^{\tau {\cal H}} \phi e^{ - \tau {\cal H}}\) with the Hamiltonian of the system ${\cal H}$.
Here, we introduced operator $\hat{\omega}$, which is defined from \( \hat{\omega}A \equiv  [A,{\cal H}] \) for any operator $A$. The right hand side of eq.~(\ref{eqn:OPM1})~is expressed as
\begin{equation}
\hat{\omega} \phi_{i} = \sum_{j}\epsilon^{(1,1)}_{ij} \phi_{j} + \phi^{(2)}_{i}.
\label{eqn:OPM2}
\end{equation}
The first term of the right hand side of eq.~(\ref{eqn:OPM2}) is regarded as ``system part", but in practice describes the part within the subspace of $\{ \phi_{i} \}$. Its coefficient is represented as $\epsilon^{(1,1)}$, while the explicit form is given later in eq.~(\ref{eqn:e11}). The second term of eq.(\ref{eqn:OPM2}) describes the ``environment part" which in fact describes the part orthogonal to the operator set $\{ \phi_{i} \}$. 
 The decomposition into the two parts in eq.~(\ref{eqn:OPM2}) is explicitly obtained by projection. 
We define the projection operator to the subspace of the operator set $\{ \phi_{i} \}$ as $\hat{P_{1}}$. 
 It is defined for Fermionic/Bosonic operators as 
 \[\hat{P_{1}} X \equiv \sum_{i,j}\{ X,\phi_{i} \}_{\pm} \left(\left( \langle \{ {\mib \phi} ,{\mib \phi} ^{\dagger} \}_{\pm} \rangle \right)^{-1} \right)_{ij} \phi_{j}. \] 
 Here, ${\mib \phi}$ is the vector notation of $\{ \phi_{i} \}$ whose $i$'th element is $\phi_{i}$. 
 Then the system part is explicitly obtained from
 \begin{equation}
 \sum_{j}\epsilon^{(1,1)}_{ij}\phi_{j} = \hat{P_{1}}(\hat{\omega} \phi_{i}).
 \label{eqn:e11}
 \end{equation}
  
 The next step is to derive the equation of motion for $\phi^{(2)}_{i}$;
 \begin{equation}
-\frac{\partial \phi^{(2)}_{i}}{\partial \tau} = \hat{\omega} \phi^{(2)}_{i}.
\end{equation}
By applying another projection operator, the right hand side becomes
\begin{equation}
{\hat \omega} \phi^{(2)}_{i} = \sum_{j} \left( \epsilon^{(2,1)}_{ij} \phi_{j} + \epsilon^{(2,2)}_{ij} \phi^{(2)}_{j} \right)+ \phi^{(3)}_{i}.
\label{eqn:OPM3}
\end{equation}
The higher order equations of motion \(\hat{\omega}\phi^{(n)}_{i}=\sum_{l=1}^{n}\sum_{j} \epsilon_{ij}^{(n,l)}\phi_{j}^{(l)}+\phi^{(n+1)} \) are obtained similarly with the matrices defined as
\begin{equation}
 \sum_{j}\epsilon^{(n,l)}_{ij} \phi^{(l)}_{j} = \hat{P}_{l} ( \hat{\omega} \phi^{(n)}_{i}). 
\end{equation}
Here, the $n$-th order operators are defined as
\begin{align}
\hat{P}_{n} X &\equiv  \{ X ,{\mib \phi} ^{(n) \dagger} \}_{ \pm } ( \langle \{ {\mib \phi} ^{(n)},{\mib \phi} ^{(n) \dagger} \}_{\pm} \rangle )^{-1} {\mib \phi} ^{(n)}, \\
{\mib \phi} ^{(n+1)} & \equiv  \prod_{l=1}^{n} (1-\hat{P}_{l}) \hat{\omega} {\mib \phi} ^{(n)}. 
\end{align}

These equations generate (imaginary) frequency-dependent correlation functions.
We introduce the notation $ \langle \langle \cdot\cdot\cdot \rangle \rangle_{i\omega_{n}} $ as the Fourier transformation of the correlation function \[ \langle \langle X ;Y^{\dagger} \rangle \rangle_{i\omega_{n}} \equiv \int_{0}^{\beta} d\tau e^{i \omega_{n} \tau} \langle -X(\tau) Y^{\dagger} \rangle. \]
We also introduce the Fourier transformation of the operators $\{\phi_{i}\}$ as 
\[ \phi_{\mib k} \equiv \sqrt{\frac{1}{N}} \sum_{i} \phi_{i} e^{i {\mib k}{\mib r} _{i}}, \]
as well as the Fourier transformation of matrices $A_{ij}$ as
\[ 
A_{\mib k} \equiv \frac{1}{N} \sum_{i} \sum_{j} A_{ij} e^{i {\mib k}({\mib r}_{i}-{\mib r} _{j})}, 
\]
where $N$ is the number of sites in the system, and ${\mib r}_{i}$ is the coordinate of the $i$'s site.
The equation of motion for the correlation function becomes
\begin{equation}
i\omega_{n} \langle \langle  \phi^{(n)}_{\mib k} ; \phi^{(n) \dagger}_{\mib k} \rangle \rangle_{i\omega_{n}} 
= 
\epsilon^{(n,n-1)}_{\mib k} + \epsilon^{(n,n)}_{\mib k} \langle \langle  \phi^{(n)}_{\mib k}; \phi^{(n) \dagger}_{\mib k} \rangle \rangle_{i\omega_{n}}
+\Sigma^{(n)}_{\mib k} \langle \langle  \phi^{(n)}_{\mib k}; \phi^{(n) \dagger}_{\mib k} \rangle \rangle_{i\omega_{n}},
\end{equation}
and thus the Green's function becomes
\begin{equation}
\langle \langle  \phi^{(n)}_{\mib k}; \phi^{(n) \dagger}_{\mib k} \rangle \rangle_{i\omega_{n}}  = \frac{\epsilon^{(n,n-1)}_{\mib k}}{i\omega_{n}-\epsilon^{(n,n)}_{\mib k}-\Sigma^{(n)}_{\mib k} }. 
\end{equation}
The self-energy $\Sigma^{(n)}$ is given as
\[ \Sigma^{(n)}_{k}=\langle \langle \phi^{(n+1)}_{k};\phi^{(n+1) \dagger}_{k}\rangle \rangle_{i\omega_{n}}. \]

If one stops the expansion at the first step, and neglects $\Sigma^{(1)}$, it obviously reproduces the Hartree-Fock Approximation (HFA). 
If one stops at the second step, and neglects $\Sigma^{(2)}$, it gives the two-pole approximation (TPA) ~\cite{Roth}. In case of the Hubbard model, for example, one can obtain the upper and lower Hubbard band at this step.
Here, in the CPM, we calculate $\Sigma^{(2)}$ to reach a  better approximation.

\subsubsection{The second procedure; DMFT}
By means of OPM in the previous section, we obtain the Green's function of the system as
\begin{equation}
G_{\mib k}(i\omega_{n})  = \dfrac{1}
                      {i\omega_{n}-\epsilon^{(1,1)}_{\mib k} 
                         -\epsilon^{(2,1)}_{\mib k}  
                                \dfrac{1}
                                {i\omega_{n}
                                     -\epsilon^{(2,2)}_{\mib k} 
                                     -\Sigma^{(2)}_{\mib k} 
                                     (i\omega_{n}) 
                                     }
                           }.
\end{equation}
In order to formulate the local approximation scheme, we reformulate the equation of motion in the previous section by the path integral.
In the following equations, we introduce the Grassmann field 
$\phi^{(n)} (\bar{\phi}^{(n)} ) $ 
corresponding to the (normalized) operators 
$\hat{\phi}^{(n)}/\sqrt{\epsilon^{(n,n-1)}}$ $(\hat{\phi}^{(n) \dagger}/\sqrt{\epsilon^{(n,n-1)}} ) $ .  
The partition function is given by
\begin{equation}
 Z=\int \prod_{i} {\cal D} \phi_{i}(\tau) {\cal D} \bar{\phi}_{i}(\tau) e^{-S},
 \label{eqn:Z}
\end{equation}
where the action has the form
\begin{multline}
 S = \int d\tau  \sum_{i,j}\bar{\phi}_{i} (\partial_{\tau} \delta_{i,j} + \epsilon^{(1,1)}_{ij}) \phi_{j} 
+ \sum_{i} \bar{\phi}^{(2)}_{i} \sqrt{ \epsilon^{(2,1)}_{i,i}} \phi_{i} 
+ \sum_{i} \bar{\phi}_{i} \sqrt {\epsilon^{(2,1)}_{i,i}} \phi^{(2)}_{i} \\
 + \sum_{i,j} \bar{\phi}^{(2)}_{i} (\partial_{\tau} + \delta_{i,j} \epsilon^{(2,2)}_{ij}) \phi^{(2)}_{j} 
+\sum_{i} \left( \bar{\phi}^{(2)}_{i}\sqrt{\epsilon^{(3,2)}} \phi^{(3)} _{i} 
+  \bar{\phi}^{(3)}_{i} \sqrt{\epsilon^{(3,2)}} \phi^{(2)}_{i} \right).
\label{eqn:action}
\end{multline}
Here, the effective action is derived so that its functional derivative makes correct equation of motion for $\phi$ and $\phi^{(2)}$.        
  If one could solve the action for $\phi^{(2)}$ exactly, one would obtain $G^{(1)}$, where $G^{(1)}$ is the full propagator for the $\phi^{(2)}$ field.
By using $G^{(1)}$, eq.(\ref{eqn:action}) is rewritten as
\begin{multline}
  S = \int d\tau  \sum_{i,j} \bar{\phi}_{i} (\partial_{\tau} \delta_{i,j} + \epsilon^{(1,1)}_{ij}) \phi_{j}
  + \sum_{i} \left(
  \bar{\phi}^{(2)}_{i} \sqrt{ \epsilon^{(2,1)}_{i,i}} \phi_{i}
 + \bar{\phi}_{i} \sqrt {\epsilon^{(2,1)}_{i,i}} \phi^{(2)}_{i} \right) \\
 + \int d\tau \int d\tau^{\prime} 
 \sum_{i,j} \bar{\phi}^{(2)}_{i}(\tau) (-G^{(1) ~-1}_{ij} (\tau-\tau^{\prime} )) \phi^{(2)}_{j}(\tau^{\prime} ).
\label{eqn:action2}  
\end{multline} 
In general, however, one cannot rigorously solve the action for $\phi^{(2)}$ fields. Here, we introduce local but imaginary-time dependent Weiss field ${\cal G}^{(1)}$ to approximate the action for the $\phi^{(2)}$ fields without the last two terms in eq.~(\ref{eqn:action}). Then, using ${\cal G}^{(1)}$, the action is rewritten as 
\begin{multline}
 S = \int d\tau  \sum_{i,j} \bar{\phi}_{i} (\partial_{\tau} \delta_{i,j} + \epsilon^{(1,1)}_{ij}) \phi_{j}
  + \sum_{i} \left( 
  \bar{ \phi_{i}} \sqrt{ \epsilon^{(2,1)}_{i,i}} \phi^{(2)}_{i} + \bar{\phi}^{(2)}_{i} \sqrt {\epsilon^{(2,1)}_{i,i}} \phi_{i} \right) \\
	+ \int d\tau \int d\tau^{\prime}  \sum_{i}
	\bar{\phi}^{(2)}_{i}(\tau) (-{ \cal G}^{(1) -1}(\tau-\tau^{\prime} )) \phi^{(2)}_{i}(\tau^{\prime} ) \\
      + \sum_{i} \left(
      \bar{\phi}^{(2)}_{i} \sqrt{\epsilon^{(3,2)}} \phi^{(3)}_{i} + \bar{\phi}^{(3)}_{i} \sqrt{\epsilon^{(3,2)}} \phi^{(2)}_{i} \right).
\label{eqn:action3}
\end{multline} 
After integrating out $\phi^{(2)}$ field, we obtain
\begin{equation}
  S = \int d\tau \int d\tau^{\prime} \sum_{i,j}  \bar {\phi}_{i}(\tau) \left(
  -{\cal G}^{(0) -1} \left(\tau-\tau^{\prime}\right)\right)_{ij} \phi_{j}(\tau^{\prime} ) 
  +\sum_{i} \left( \bar{\phi}^{(2)}_{i} \sqrt{\epsilon^{(3,2)}} \phi^{(3)}_{i} + \bar{\phi}^{(3)}_{i} \sqrt{\epsilon^{(3,2)}} \phi^{(2)}_{i} \right)
\label{eqn:action4}
\end{equation}
Here, the `Weiss field' ${\cal G}^{(0)}$ is given as
\begin{equation}
{\cal G}^{(0) -1}_{ij} (\tau-\tau^{\prime}) = (- \partial_{\tau} \delta_{i,j} - \epsilon^{(1,1)}_{ij}) \delta(\tau-\tau^{\prime})
 -\sqrt{\epsilon^{(2,1)} }{\cal G}^{(1)}(\tau-\tau^{\prime} ) \sqrt{\epsilon^{(2,1)}}.
 \label{eqn:g0k}
\end{equation}
With the Weiss field, we can now solve $\Sigma^{(2)}$.
As was shown in the last section, $\Sigma^{(2)}$ has the form
\begin{equation}
\Sigma^{(2)}=\langle \langle \phi^{(3)};\phi^{(3) \dagger} \rangle \rangle_{i\omega_{n}},
\label{eqn:sgm2}
\end{equation}
where the field $\phi^{(3)}$ can be expressed with the original $\phi$ field. Since we have the Weiss field for $\phi$ field, we can solve the `second-order self-energy' in eq.(\ref{eqn:sgm2}) with some proper method. In this paper, we adopt the iterative perturbation theory (IPT) as we discuss below.

The whole iteration procedure is explicitly given below:\\
\noindent{[1] IPT solution to $\Sigma^{(2)}$}\\
Evaluate the self-energy $\Sigma^{(2)}$ as 
\begin{equation}
\Sigma^{(2)}=\dfrac{1}{N} \sum_{k} \dfrac{1}{\epsilon^{(2,1)}_{k}}\langle \langle \phi^{(3)}_{k};\phi^{(3) \dagger}_{k} \rangle \rangle_{i\omega_{n}}.
\label{eqn:ipt1}
\end{equation}
The operator $\phi^{(3)}$ is evaluated by Wick expansion. For $\phi^{(3)}$ expressed in the form
\(
 \phi^{(3)}_{\mbox{$k$}} 
= 
\dfrac{1}{N^{2}} \sum_{k^{\prime},q}\Gamma_{\mbox{$k,k^{\prime},q$}} \phi^{\dagger}_{k^{\prime}+q} \phi_{k^{\prime}} \phi_{k-q},
\)
we can expand $\Sigma^{(2)}$ by IPT as
\begin{equation}
\Sigma^{(2)}=
\dfrac{1}{N^{3}} 
	\sum_{k,k^{\prime},q} 
		\dfrac{1}{\epsilon^{(2,1)}_{k}} \Gamma_{\mbox{$k,k^{\prime},q$}}
			{\cal G}^{(0)}_{\mbox{$k^{\prime}$}}(-\tau){\cal G}^{(0)}_{\mbox{$k^{\prime}+q$}}(\tau) {\cal G}^{(0)}_{\mbox{$k-q$}}(\tau).
\label{eqn:ipt2}
\end{equation}
For the initial condition in the first iteration, one may use the Green's function for the free particle for ${\cal G}^{(0)}$.\\
{[2] self-consistency equation}\\
The local self-consistency condition is given as
\begin{equation}
G^{(1)}_{loc}(i\omega_{n}) = \dfrac{1}{N} \sum_{\mbox{$k$}}  \dfrac{1}{i\omega_{n}-\epsilon_{\mbox{$k$}} ^{(2,2)}-\Sigma^{(2)}(i\omega_{n})}.
\label{eqn:sce}
\end{equation}
{[3] Dyson equation for ${\cal G}^{(1)}$}\\
The Dyson equation for the field $\phi^{(2)}$ is given as
\begin{equation}
{\cal G}^{(1) -1}(i\omega_{n})=G_{loc}^{(1) -1}(i\omega_{n}) + \Sigma^{(2)}(i\omega_{n}).
\label{eqn:dyson}
\end{equation}
{[4] Equation for `Weiss field' ${\cal G}_{\mbox{$k$}}$}\\
From the eq.(\ref{eqn:g0k}), we can calculate the momentum dependent Weiss field for the field $\phi_{k}$ as
\begin{equation}
  {\cal G}^{(0) -1}_{\mbox{$k$}} (i\omega_{n}) = i\omega_{n} - \epsilon^{(1,1)}_{\mbox{$k$}}  -\sqrt{\epsilon^{(2,1)}_{\mib k}} {\cal G}^{(1)}(i\omega_{n}) \sqrt{\epsilon^{(2,1)}_{\mib k}}.
\label{eqn:weiss} 
\end{equation}
After Fourier transform from ${\cal G}^{(0)}_{k}(i\omega_{n})$ to ${\cal G}^{(0)}_{k}(\tau)$,  ${\cal G}^{(0)}_{k}(\tau)$ is substituted to eq.(\ref{eqn:ipt2}) and we repeat the procedure from [1].
The procedure [1]-[4] constitutes the iteration loop.
The procedure is repeated until we reach the convergence. After the convergence, the Green function is obtained as
\begin{equation}
G_{k} (i\omega_{n}) = \dfrac{1}
                       {i\omega_{n}-\epsilon^{(1,1)}_{k} 
                            -\dfrac{\epsilon^{(2,1)}_{k}}
                             {i\omega_{n}- \epsilon^{(2,2)}_{k} 
                             -\Sigma^{(2)}(i\omega_{n})}
                             }.
\label{eqn:finalGF}
\end{equation}

\subsection{Application to the extended Hubbard Model}
We now apply this method to the extended Hubbard model defined by eq.(\ref{eqn:EHM}). Here, we introduce two methods in analyzing the model.

\subsubsection{Standard approach} 
We first apply the OPM procedure in a straightforward manner. Now we take $c_{i \sigma}$ as the first set of the operators $\phi_{i}$ in eq.~(\ref{eqn:OPM1}).

The equation of motion eq.~(\ref{eqn:OPM1}) reads,
\begin{equation}
\hat{\omega} c_{i\sigma} = \sum_{j} \epsilon^{(1,1)}_{ij} c_{j\sigma} + \psi_{i\sigma}.
\end{equation}
The matrix element is given as
\[\epsilon^{(1,1)}_{ij}=-t_{ij} + \langle \Phi_{i\sigma}\rangle \delta_{i,j} - X_{ij},\]
where \( \Phi_{i\sigma} = U n_{i-\sigma} + \sum_{j}V_{i,j} n_{j} \) and
 \( X_{i,j}=V_{ij} \langle \rho_{j,i,\sigma}\rangle  \) are direct and exchange fields with the notation
\( \rho_{i,j,\sigma} \equiv c_{i\sigma}^{\dagger} c_{j\sigma}\),  
\(n_{i} \equiv n_{i\uparrow} + n_{i\downarrow} \).
We also define the following notations;
\(\delta \Phi_{i\sigma}= \Phi_{i\sigma} - \langle \Phi_{i\sigma} \rangle \), 
\( \Phi_{i\sigma}^{V}=\sum_{j} V_{ij} n_{j}. \)
\( \delta \Phi_{i\sigma}^{V}=\Phi_{i\sigma}^{V} - \langle \Phi_{i\sigma}^{V} \rangle. \)
The second operator $\psi_{i\sigma}$ is obtained as
\(\psi_{i\sigma}= \delta \Phi_{i\sigma} c_{i\sigma} + V_{i,j}\langle \rho_{i,j,\sigma}\rangle c_{j\sigma}. \)
    
We can now proceed to the equation of motion for the second set of operators,
which is formally written as
\begin{equation}
\hat{\omega} \psi_{i} = \sum_{j} \left(
\epsilon^{(2,1)}_{ij} c_{j} + \epsilon^{(2,2)}_{ij} \psi_{j} \right)+ \mit {R_{i}},
\label{eqn:strt2}
\end{equation}
where the matrices are written as follows;
\[
\epsilon^{(2,1)}_{ij}
=\langle \{ \psi_{i},\psi^{\dagger}_{j} \} \rangle 
=\langle  ( \delta \Phi_{i})^{2} \rangle \delta_{i,j}- | V_{il} \langle \rho_{i,l}\rangle |^{2},
\]
\[
\epsilon^{(2,2)}_{ij}
= -\tilde{t}_{ij}-\tilde{X}_{ij} + 
                         ( \langle \Phi_{i\sigma}^{\rm eff} \rangle -\delta \mu_{i})\delta_{i,j},
\]
\begin{equation}
\begin{split}
\tilde{t}_{ij}
&= \left( 
U^{2} \langle 
		\frac{1}{4}\delta n_{i}\delta n_{l} 
        + \mbox{ 
          		\boldmath{$S$}$_{i}$ \boldmath{$S$}$_{l}$
          		}
        -\Delta^{p \dagger}_{i} \Delta^{p}_{l} 
      \rangle + 
\langle \delta \Phi^{V}_{i} \delta \Phi^{V}_{l} \rangle 
    \right) t_{il}\\          
&\qquad \qquad \times \left( ( \epsilon^{(2,1)} )^{-1}
                       \right)_{lj},\nonumber
\end{split}
\end{equation}
\[
\delta \mu_{i}= \left( - U^{2} \sum_{j} t_{ij} \langle \rho_{i,j,-\sigma} \rangle
-\sum_{j,\sigma^{\prime}} V^{2}_{ij} t_{jl}\langle \rho_{j,l,\sigma^{\prime}}\rangle \right)(1-2\langle n_{i\sigma}\rangle) \left( \epsilon^{(2,1)} \right)^{-1},
\] 
\[\tilde{X}_{ij} = V_{ij} \langle \rho_{ji\sigma} \rangle
                     \langle 2 \delta \Phi_{i\sigma} \delta \Phi_{j\sigma} 
                              - (\delta \Phi_{i\sigma})^{2}
                              - (\delta \Phi_{j\sigma})^{2} \rangle,   \]
\begin{equation}
\langle \Phi_{i\sigma}^{\rm eff} \rangle
= \dfrac{ \langle \Phi_{i\sigma} (\delta \Phi_{i\sigma})^{2} \rangle}
            {\langle ( \delta \Phi_{i\sigma} )^{2} \rangle}.
\end{equation}
Here, $\mbox{\boldmath{$S$}}_{i}$ is the spin operator on the site $i$, defined as \( \mbox{\boldmath{$S$}$_{i}$} \equiv \frac{1}{2} c^{\dagger}_{i\alpha} \mbox{\boldmath{$\sigma$}}_{\alpha,\beta}c_{j\beta}$ with Pauli matrices $\mbox{\boldmath{$\sigma$}}\).
The singlet pair operator on the site $i$, $\Delta^{p}_{i}$ is defined as $\Delta^{p}_{i} \equiv c_{i\uparrow}c_{i\downarrow}$.
The motion of the interacting charges is reflected in $\delta \mu$. 
The term with $\Phi^{\rm eff}$ comes from the interaction part. It differs from $\langle \Phi \rangle$ because of the Pauli principle. ({\it e.g.} $n_{i-\sigma}n_{i-\sigma}c_{i\sigma}= n_{i-\sigma}c_{i\sigma}$)

The remaining operator, ( the third operator of the right hand side of the eq.~(\ref{eqn:strt2}) ) $R$ reflects the fluctuation of $\delta \Phi$. 
 It is given as
\begin{multline}
 R_{i\sigma} 
	= U  \sum_{j}\left(
		(-t_{ij}) \left( \rho_{i,j,-\sigma}-\rho_{j,i,-\sigma}\right) 
		c_{i\sigma} - t_{ij}\delta n_{i,-\sigma}c_{j\sigma} 
		\right) \nonumber \\
	+    
	 \sum_{j\sigma^{\prime}} V_{ij} \left( 
	\sum_{l}(-t_{jl}) \left( \rho_{j,l,\sigma^{\prime}}- \rho_{l,j,\sigma^{\prime}}\right)c_{i\sigma} -t_{il} \delta n_{j\sigma^{\prime}}c_{l\sigma} 
	\right)\\          
	\qquad+ \sum_{j} ( \tilde{t}_{ij}+\delta \mu \delta_{ij} ) \psi_{j\sigma}
	+R^{\prime}_{i\sigma}.
\end{multline}
Here, $R^{\prime}_{i\sigma}$ contains higher order operators as
\[
  R^{\prime}_{i\sigma}=
  \delta {\cal Q}_{i\sigma}c_{i\sigma}
 \]
 with $\delta {\cal Q}_{i\sigma}$ defined as \(\delta {\cal Q}_{i\sigma}\equiv (\delta \Phi_{i\sigma})^{2} - \dfrac{\langle (\delta \Phi_{i\sigma})^{3} \rangle}{\langle (\delta \Phi_{i\sigma})^{2}\rangle}\delta \Phi_{i\sigma}. \)
The correlation functions of this operator give a convolution of Green's function with charge/spin correlation functions of more than three terms such as $\langle \langle \delta n_{i} \delta n_{j} \delta n_{l} \rangle \rangle$ in which indices $i$, $j$ and $l$ are all different. 
Assuming that these higher order fluctuations are small, we neglect this operator $R^{\prime}_{i\sigma}$.

Here, in the matrix elements, one finds equal-time correlation functions like $\langle (\delta \Phi)^{2} \rangle$ that cannot be uniquely determined by the single-particle Green's function. There are several ways to determine these correlation functions.\\
(1) The second-order perturbation theory: One way is to make use of one of various perturbative methods. Among them, we adopt the second order perturbation theory. 
Here, we do not start from the free-particle Green function, but use obtained (renormalized) Green function and take up to the second order diagrams to include correlation effects. 
Although it is one of the simplest approximations, it is favorable for our purpose here since it conserves the local summation rule~\cite{TPSC};
\begin{equation}
\frac{1}{N \beta}\sum_{\omega_{n}}\sum_{q} 
\left(
\chi_{q}^{({\rm ch})}(i\omega_{n})
+ \chi_{q}^{({\rm sp})}(i\omega_{n}) 
\right)
	= \langle n \rangle (2-\langle n \rangle).
\end{equation}
Here, $\chi_{q}^{({\rm ch})}(i\omega_{n})\equiv \langle \langle n_{q};n_{-q}\rangle \rangle_{i\omega_{n}}$ and $\chi_{q}^{({\rm sp})}(i\omega_{n}) \equiv \langle \langle (n_{q \uparrow}-n_{q \downarrow});(n_{-q \uparrow}-n_{-q \downarrow})\rangle \rangle_{i\omega_{n}}$ are the charge and spin correlation functions, respectively.
By the second order perturbation theory, they are derived as
\begin{equation}
\begin{split}
\chi_{q}^{({\rm ch})}(i\omega_{n})
&=2 \{
\chi_{q}^{0}(i\omega_{n}) 
+ \chi_{q}^{0}(i\omega_{n})(U+2V_{q}) \chi_{q}^{0}(i\omega_{n})\\
&\quad- \frac{1}{N^{2}}\sum_{k, k^{\prime}}
             \chi^{0}_{k;q}(i\omega_{n})V_{k-k^{\prime}}
             \chi^{0}_{k^{\prime};q}(i\omega_{n})    
             \},
\end{split}
\end{equation}
\begin{equation}
\begin{split}
\chi_{q}^{({\rm sp})}(i\omega_{n})
&=2\{\chi_{q}^{0}(i\omega_{n}) 
           - \chi_{q}^{0}(i\omega_{n}) U \chi_{q}^{0}(i\omega_{n})  \\
&\quad - \frac{1}{N^{2}}\sum_{k, k^{\prime}}
             \chi^{0}_{k;q}(i\omega_{n})V_{k-k^{\prime}}
             \chi^{0}_{k^{\prime};q}(i\omega_{n}) \}.          
\end{split}
\end{equation}
Here, $\chi_{q}^{0}(i\omega_{n})$ and $\chi^{0}_{k;q}(i\omega_{n})$ are zeroth order susceptibility functions defined as
\begin{equation}
\chi_{q}^{0}(i\omega_{n})=\frac{1}{N\beta}\sum_{i\omega_{l}}\sum_{k} 
G_{k+q}(i\omega_{l}+i\omega_{n})G_{k}(i\omega_{l})
\end{equation}
\begin{equation}
\chi_{k;q}^{0}(i\omega_{n})
=\frac{1}{\beta} \sum_{i\omega_{l}}
    G_{k+q}(i\omega_{l}+i\omega_{n})G_{k}(i\omega_{l}),
\end{equation}
where $G_{k}(i\omega_{n})$ is determined from eq.(\ref{eqn:finalGF}).
On the other hand, it may underestimate the local correlation effects especially in the case of strong coupling. \\
(2) The exact diagonalization (ED) of small cluster:
We can also take a completely different approach by simply replacing the correlation functions with the result of exact diagonalization (ED) of a small cluster. Adopting ED, one can take account of the strong local correlation effects. Here we emphasize that we only need {\it short-ranged} equal time correlation functions that are not likely to severely suffer from finite size effects. 
In order to treat the symmetry-broken state in the ED calculation, we divide the lattice into two sublattices, $A$ and $B$, and apply small symmetry-breaking field $h$ as follows;
\[
{\cal H}_{\rm eff}={\cal H}_{\rm EHM} - h \left( \sum_{i\in A}n_{i}-\sum_{j \in B}n_{j} \right).
\]
Here, the first term ${\cal H}_{\rm EHM}$ is the original Hamiltonian of the extended Hubbard model given in eq.~(\ref{eqn:EHM}), while the second and third term indicate the symmetry breaking field. The field $h$ is set $h=0.001t$.
The shortcoming of this approach is that the correlation functions are determined independently of the lattice Green's function, and are not fully self-consistent.

We adopt the former approach to derive the charge ordering transition where the consistency to the Green's function is particularly important. On the other hand, we adopt the latter approach to analyze the detailed properties for the charge-ordered phase.

\subsubsection{Formulation based on the CDW basis}
In order to further analyze the properties of the charge-ordered phase, we introduce a different approach using different basis representation. Although in principle, the operator projection theory does not depend on the choice of basis, in actual calculations, however, details of the results do depend on the chosen basis set due to the truncation of the operator expansion at a finite step.
Because of the truncation, it is better to employ the basis function which is the eignfunction of the same symmetry as that of the correct phase. For example, in the charge-disordered state, the basis function of the free-particle Hamiltonian gives faster convergence and gives better results at a fixed level of the truncation. This is the case in the previous subsection, namely, the standard approach. On the other hand, if the system is in the charge ordered phase, the basis function of charge density wave (CDW) mean-field solution is expected to give better results. 
Here, we introduce an approximate formulation which is valid under large polarization. The idea is to solve the equation of motion for the operators represented by the eignfunctions of the mean-field Hamiltonian for the charge order.

In the following equations, we consider the square lattice as an example.
Hence, at quarter filling, two sublattices A and B are enough to discuss possible charge ordering. 
The dispersion in the two-dimensional square lattice is denoted as \(-t_{\mib k}\equiv  -2t(\cos(k_{x}) + \cos(k_{y})) \), and electron density per site as $\langle n_{A} \rangle$ and $\langle n_{B} \rangle$, for each sublattice. We also use the notation \( \langle n \rangle \equiv \frac{1}{2}( \langle n_{A} \rangle + \langle n_{B} \rangle ) \)
We take new basis as
\begin{subequations}
\begin{eqnarray}
 a_{\mib k}&=&u_{\mib k} c^{A}_{\mib k} + v_{\mib k}c_{\mib k}^{B}  \\
 b_{\mib k}&=&-v_{\mib k} c^{A}_{\mib k} + u_{\mib k} c_{\mib k}^{B}.
\end{eqnarray}
\label{eqn:eigenvec}
\end{subequations}
Here, $c^{R}_{\mib k} ( c^{R \dagger}_{\mib k} )$ is the annihilation (creation) operator of the Bloch state at sublattice $R$ (\(R=A \ or \  B \)), with momentum $\mbox{ \boldmath $k$ } $. Here, and in the following equations in this section, $\mbox{ \boldmath $k$ } $ refers to the wavevector inside the reduced Broullin zone (RBZ). We fix $u_{\mib k}$ and $v_{\mib k}$ as the eigenvector of the mean-field Hamiltonian matrix $\epsilon^{\rm HF}$. 
The Hubbard term $U\sum_{i}n_{i\uparrow}n_{i\downarrow}$ is treated in later calculations. 
We define the CDW order parameter $\Delta_{\rm CDW}$ as
\begin{equation}
\Delta_{\rm CDW}\equiv \frac{1}{2} \times 4V (\langle n_{A} \rangle -\langle n_{B} \rangle ). 
\label{eqn:CDWgap}
\end{equation}
The order parameter is to be determined in a self-consistent manner.
The mean-field Hamiltonian becomes
\begin{eqnarray}
 \epsilon_{\mib k}^{\scriptsize \mbox{HF}} = \left(
 \begin{array}{cc}
 4V\langle n \rangle -\Delta_{\rm CDW} & -t_{\mib k} \\
 -t_{\mib k} & 4V\langle n \rangle + \Delta_{\rm CDW}.
 \end{array}
 \right).
\end{eqnarray}
Hence $u_{\mib k}$ and $v_{\mib k}$ is obtained by the following equations:
\begin{equation}
 \epsilon_{\mib k}^{\scriptsize \mbox{HF}} \left(
 \begin{array}{cc}
 u_{\mib k} & v_{\mib k} \\
 -v_{\mib k} & u_{\mib k}
 \end{array}
 \right) 
 =\left(
 \begin{array}{cc}
 u_{\mib k} & v_{\mib k} \\
 -v_{\mib k} & u_{\mib k}
 \end{array}
 \right) \left(
 \begin{array}{cc}
 \lambda^{A}_{\mib k} & 0 \\
 0 & \lambda^{B}_{\mib k}
 \end{array}
 \right).
\end{equation}
Here, \(\lambda^{A}_{\mib k}= 4V\langle n \rangle -\lambda_{\mib k} \) and 
 \( \lambda^{B}_{\mib k}= 4V\langle n \rangle + \lambda _{\mib k} \) are eignvalues of the mean-field Hamiltonian with   
\begin{equation}
\lambda_{\mib k}= \sqrt{ \Delta_{\rm CDW}^{2} + t_{\mbox{ \boldmath $k$ }}^{2} }. 
\label{eqn:eignvalue}  
\end{equation}
The eigenvectors are determined from
\begin{subequations}
\begin{eqnarray}
 u_{\mib k} 
      &=& \sqrt{ 
          \frac{ \lambda_{\mbox{ \boldmath $k$ }} + \Delta_{\rm CDW}  }
             {2 \lambda_{\mib k}}
        }, \\
 v_{\mib k} 
      &=& \sqrt{ 
          \frac{ \lambda_{\mbox{ \boldmath $k$ }} -\Delta_{\rm CDW}  }
                {2 \lambda_{\mib k}}
         }. 
\end{eqnarray}
\label{eqn:eigenvec2}
\end{subequations}
We perform inverse Fourier-transformation of eq.(\ref{eqn:eigenvec}) to obtain site representation as
\begin{subequations}
\begin{eqnarray}
 a_{i}=\dfrac{1}{\sqrt{N}} \sum_{\mib k} a_{\mib k} e^{i \mbox{ \boldmath $k$ }  \mbox{\boldmath $r_{i}$} } \ \ i\in A, \\
 b_{j}=\dfrac{1}{\sqrt{N}} \sum_{\mib k} b_{\mib k} e^{i \mbox{ \boldmath $k$ }  \mbox{\boldmath $r_{j}$} } \ \ j\in B.
\end{eqnarray} 
\label{eqn:eigenvec3}
\end{subequations}
The last expression also satisfies the anticommutation relation;
\begin{subequations}
\begin{eqnarray}
\{ a_{i},a_{i^{\prime}}^{\dagger}  \} = \delta_{i,i^{\prime}}, \\ 
\{ b_{j},b_{j^{\prime}}^{\dagger}  \}= \delta_{j,j^{\prime}}, 
\end{eqnarray}
\end{subequations}
and any other pairs than these anticommute.
Note that in spite of the site index, they have nonzero amplitudes over many lattice sites as
\begin{subequations}
 \begin{eqnarray}
 a_{i}=\sum_{i^{\prime}} u_{i,i^{\prime}}c^{A}_{i^{\prime}} + \sum_{j} v_{i,j} c^{B}_{j}, \\
 b_{j}=- \sum_{i} v_{j,i}c^{A}_{i} + \sum_{j^{\prime}}u_{j,j^{\prime}} c^{B}_{j^{\prime}},
 \end{eqnarray} 
 \label{eqn:eigenop}   
\end{subequations}
where $u_{i,j}=u({\mib r}_{i}-{\mib r}_{j})$ and  $v_{i,j}=v({\mib r}_{i}-{\mib r}_{j})$ indicate the Fourier transformation of $u_{\mib k}$ and $v_{\mib k}$. 
 The Hamiltonian can be written with the new operators. 
We introduce the following notations; 
 \[
 s_{i,x} =\left \{
 \begin{array}{ll}
 \displaystyle{ u_{i,x}} & \qquad x \in A, \\
 \displaystyle{-v_{i,x}} & \qquad x \in B,
 \end{array} \right. 
 \]
 \[
 q_{j,x} =\left \{
 \begin{array}{ll}
 \displaystyle{ v_{j,x}} & \qquad x \in A, \\
 \displaystyle{ u_{j,x}} & \qquad x \in B,
 \end{array} \right. 
 \]
 \[
 d_{x,\sigma} =\left \{
 \begin{array}{ll}
 \displaystyle{ a_{x,\sigma} } & \qquad x \in A, \\
 \displaystyle{ b_{x,\sigma} } & \qquad x \in B.
 \end{array} \right. 
 \]
  We write the Hamiltonian as
 \begin{equation}
 {\cal H}={\cal H}_{\scriptsize \mbox{mf}} + {\cal H}_{U} + {\cal H}^{\prime}. 
 \end {equation}
 Here, ${\cal H}_{\scriptsize \mbox{mf}}$ is the mean-field terms;
 \begin{equation}
 {\cal H}_{\scriptsize \mbox{mf}} = -\sum_{(i,j)} t_{i,j}c_{i \sigma}^{\dagger} c_{j \sigma} + \sum_{i} \sum_{j} V_{i,j} \langle n_{j}\rangle  n_{i} 
 -\frac{1}{2} \sum_{i} \sum_{j} V_{i,j} \langle n_{i}\rangle \langle n_{j}\rangle
 \label{eqn:Hmf}
 \end{equation}
 and ${\cal H}_{U}$ is the Hubbard interaction terms;
\begin{equation}
 {\cal H}_{U} = \sum_{i \in A} U n^{A}_{i\uparrow} n^{A}_{i\downarrow}
                + \sum_{j \in B} U n^{B}_{j\uparrow} n^{B}_{j\downarrow}.
\label{eqn:Hu}
\end{equation}
The rest part of the mean field Hamiltonian is given by
\begin{equation}
 {\cal H}^{\prime} = \sum_{(i,j)} V_{i,j} \delta n^{A}_{i} \delta n^{B}_{j}.
\label{eqn:Hprime}
\end{equation}
 Then, ${\cal H}_{\scriptsize \mbox{mf}}$ is rewritten as
 \begin{equation}
 {\cal H}_{\scriptsize \mbox{mf}}=\sum_{i,i^{\prime} \in A} \lambda^{A}_{i,i^{\prime}} a_{i}^{\dagger} a_{i^{\prime}} + \sum_{j,j^{\prime} \in B} \lambda^{B}_{j,j^{\prime}} b_{j}^{\dagger} b_{j^{\prime}}  + E_{\scriptsize \mbox{mf}}. 
 \end{equation}
 Here, $\lambda^{A}_{in}$ and $\lambda^{B}_{jm}$ is the Fourier transform of $\lambda^{A}_{\mib k}$ and $\lambda^{B}_{\mib k}$ within each sublattice, respectively, and $E_{ \scriptsize \mbox{mf} }$ is the mean field energy;
 \[ 
 E_{ \scriptsize \mbox{mf} } \equiv 2 V \langle n \rangle N_{e}+ \frac{\Delta_{\rm CDW}^{2}}{8V}N, 
 \]
 with $N_{e}$ being the total electron number.
 Whereas ${\cal H}^{\prime}$ and ${\cal H}_{U}$ are written by substituting 
\begin{subequations}
\begin{eqnarray}
n_{i \sigma} = \sum_{x,y} s_{i,x} s_{i,y} d^{\dagger}_{x \sigma} d_{y \sigma} \\n_{j \sigma} = \sum_{x,y} q_{j,x} q_{j,y} d^{\dagger}_{x \sigma} d_{y \sigma}.
\end{eqnarray}
\end{subequations}
into the original number operators $n^{R}_{x\sigma}$ in eqs.~(\ref{eqn:Hu}),(\ref{eqn:Hprime}).
 
 In general case, the above-mentioned procedure yields complication. However, it becomes simple in the largely disproportionated case.
 In the following of this section, we consider such case.
 We assume the CDW order parameter $\Delta_{\rm CDW}$ to be considerably large so that 
 we define an expansion parameter $p$ as
 \begin{equation}
 p \equiv \frac{t}{2 \Delta_{\rm CDW}}.
 \label{eqn:defof_p}
 \end{equation} 
 Then $u_{\mib k}$ and $v_{\mib k}$ in eq.(\ref{eqn:eigenvec2}) are expanded in power of $p$ up to $p^{2}$ as
\begin{subequations}
 \begin{eqnarray}
 u_{\mib k}&=&1-\frac{t_{\mib k}^{2}}{4 \Delta_{\rm CDW}^{2}} + {\mit O}( p^{4} ) \\
 v_{\mib k}&=&\frac{t_{\mib k}}{2 \Delta_{\rm CDW}} + {\mit O}( p^{3} ).
 \end{eqnarray}
 \end{subequations}
 Up to this order, their transformations to the real-space representation have nonzero values only within the nearest neighbor sites.
 We introduce the notation $p_{ij}$ as
 \[
 p_{i,j} =\left \{
 \begin{array}{ll}
 \displaystyle{p} & \qquad (i,j) \quad \mbox{ for the nearest neighbor pair}  \\
 \displaystyle{0} & \qquad \mbox{otherwise} \quad.
 \end{array} \right.   
 \]
 Then $u$ and $v$ become
 \begin{subequations}
 \begin{eqnarray}
 u_{i,i^{\prime}}=\delta_{i,i^{\prime}} + {\mit O}( p^{2}), \\
 v_{i,j}= p_{i,j} + {\mit O}( p^{3}).
 \end{eqnarray}
 \end{subequations} 
 In the above equation, $i$ and $i^{\prime}$ are on the same sublattice, whereas $j$ is on the other. The original operators $c^{A}_{i}$ and $c^{B}_{j}$ are expressed by $a_{i}$ and $b_{j}$ as
 \begin{subequations}
 \begin{eqnarray}
 c^{A}_{i}= f (a_{i} - \sum_{j} p_{i,j} b_{j}), \\
 c^{B}_{j}= f (\sum_{i} p_{j,i}a_{i} + b_{j} ),
 \end{eqnarray} 
 \end{subequations}
 with $f$ being the normalization factor \( f= \frac{1}{\sqrt{1+4p^{2}}} \).
 The number operators are rewritten as
 \begin{subequations}
 \begin{eqnarray}
 n^{A}_{i\sigma}= f^{2} (\eta^{a}_{i\sigma} - \sum_{j} p_{i,j} ( a^{\dagger}_{i\sigma}b_{j\sigma} + b^{\dagger}_{j\sigma} a_{i\sigma})), \\
 n^{B}_{j\sigma}= f^{2} (\eta^{b}_{j\sigma} + \sum_{i} p_{i,j} ( a^{\dagger}_{i\sigma}b_{j\sigma} + b^{\dagger}_{j\sigma} a_{i\sigma})), 
 \end{eqnarray} 
 \end{subequations}
 where $ \eta^{a}_{i\sigma} \equiv a_{i\sigma}^{\dagger}a_{i\sigma}$, and $ \eta^{b}_{j\sigma} \equiv b_{j\sigma}^{\dagger}b_{j\sigma}$.
 At this point it is clear that the factor $f$ can be absorbed in the interaction parameter $U$ and $V$ as \( \tilde{U} \equiv U f^{4} \) and \( \tilde{V} \equiv V f^{4} \). The Hubbard interaction term contains `off-site' terms that describe the interaction with states centered at different sites.

 We write down the kinetic equation for the operator $a_{i}$ up to the first order of $p$ as
\begin{equation}
\begin{split}
\hat{\omega} a_{i\sigma}
&=\sum_{i^{\prime}} \lambda_{i,i^{\prime}} a_{i^{\prime} \sigma}
        +\Delta \Phi_{i\sigma} a_{i\sigma}
        -\sum_{j} p_{ij}( \Delta \Phi_{i\sigma}-\Delta \Phi_{j\sigma} ) b_{j\sigma}\\
&=\sum_{i^{\prime}}\lambda_{i,i^{\prime}} a_{i^{\prime}\sigma}
      	    +\tilde{U} \langle \eta^{a}_{i\-\sigma}\rangle a_{i\sigma}
      	    +\delta \Phi_{i}a_{i\sigma} \\
&\quad -\sum_{j}p_{ij}\tilde{U}(\langle \eta^{a}_{i -\sigma}\rangle
        		-\langle \eta^{b}_{j -\sigma} \rangle ) b_{j\sigma} 
        	-\sum_{j}p_{ij}(\delta \Phi_{i}-\delta \Phi_{j})b_{j\sigma} \\
&= 
      	\sum_{i^{\prime}}\lambda_{i,i^{\prime}} a_{i^{\prime}\sigma}
      	    +\tilde{U} \langle \eta^{a}_{i\-\sigma} \rangle a_{i\sigma}\\
&\quad      	    
      	    -\sum_{j}p_{ij}\tilde{U}(\langle \eta^{a}_{i -\sigma}\rangle
        		-\langle \eta^{b}_{j -\sigma} \rangle ) b_{j\sigma} 
        		+f_{i\sigma}
      	    \label{eqn:aomg}
\end{split}
\end{equation}
with the "second" operator $f_{i\sigma}$, defined as
\begin{equation}
 f_{i \sigma} 
 \equiv 
 \delta \Phi_{i\sigma}a_{i\sigma}
        	-\sum_{j}p_{ij}(\delta \Phi_{i\sigma}-\delta \Phi_{j\sigma}) b_{j\sigma}. 
 \end{equation}
 Here, we have introduced the notation $\Delta \Phi_{i \sigma}$, defined as
 \( \Delta \Phi_{i\sigma} \equiv \Phi_{i\sigma}-\sum_{j}V_{ij}\langle n_{j} \rangle. \) Other notations are the same as in the previous subsection.
In the derivation of eq.~(\ref{eqn:aomg}), we have used the fact that terms which are off-diagonal in `$a$' and `$b$' bands such as $\langle a_{i\sigma}^{\dagger}b_{j} \rangle$ are of order $p$ or higher, and terms like $p \langle a_{i\sigma}^{\dagger}b_{j} \rangle$ can be neglected when we consider the order up to $O(p)$.
The equation for the fields $b_{j\sigma}$'s is also obtained in the same way.
\begin{equation}
\begin{split}
\hat{\omega} b_{j\sigma}
&=\sum_{m} \lambda_{j,m} b_{m\sigma}
        +\Delta \Phi_{j\sigma}b_{j\sigma}
        -\sum_{i} p_{ji}(\Delta \Phi_{i\sigma}-\Delta \Phi_{j\sigma})a_{i\sigma} \\
      &= 
      	\sum_{m} \lambda_{j,m} b_{m\sigma}
      	    +\tilde{U} \langle \eta^{b}_{j-\sigma} \rangle b_{j\sigma}\\
      	    &\quad-\sum_{i}p_{ji}(\langle \tilde{U} \eta_{i-\sigma}^{a}\rangle
      	              -\langle \tilde{U} \eta_{j-\sigma}^{b}\rangle )a_{i\sigma} 
        	+g_{j\sigma},
        	\end{split}
\end{equation}
 with the corresponding ``second" operator $g_{j\sigma}$,
\begin{equation}
 g_{j \sigma} 
 \equiv 
 \delta \Phi_{j\sigma}b_{j\sigma}
        	-\sum_{i}p_{ji}(\delta \Phi_{i\sigma}-\delta \Phi_{j\sigma})a_{i\sigma}. 
 \end{equation}

 The equation of motion for the field $f_{i \sigma}$ becomes
 \begin{equation}
\begin{split}
 \hat{\omega} f_{i\sigma}
 &=
 \hat{\omega}(\Phi_{i\sigma}) a_{i\sigma}
  + \delta \Phi_{i\sigma}(\hat{\omega}a_{i\sigma})\\
  & \quad -\sum_{j}p_{ij}( \hat{\omega}(\Phi_{i\sigma}-\Phi_{j\sigma}))b_{j\sigma}
  -\sum_{j}p_{ij}( \delta \Phi_{i\sigma} - \delta \Phi_{j\sigma}) \hat{\omega}b_{j\sigma}
  \\
 &=
  \hat{\omega}(\Phi_{i\sigma}) a_{i\sigma}
  + \delta \Phi_{i\sigma}(\sum_{n}\lambda^{A}_{in}a_{n\sigma}+\Delta \Phi_{i\sigma}a_{i\sigma})
  \\ 
  & \quad -\sum_{j}p_{ij}( \hat{\omega}(\Phi_{i\sigma}-\Phi_{j\sigma}) )b_{j\sigma}
  \\
  & \qquad -\sum_{j}p_{ij}( \delta \Phi_{i\sigma} - \delta \Phi_{j\sigma}) 
               (\sum_{m}\lambda^{B}_{jm}b_{m\sigma} + \Delta \Phi_{j\sigma}b_{j\sigma})
               \\
  &=\sum_{n}\epsilon^{(aa) (2,1)}_{in}a_{n\sigma}+\sum_{j}\epsilon^{(ab) (2,1)}_{ij}b_{j\sigma}\\
 &\quad +\sum_{n}\epsilon^{(aa) (2,2)}_{in}f_{n\sigma}+\sum_{j}\epsilon^{(ab) (2,2)}_{ij}g_{j\sigma}
 +L_{i\sigma},
 \end{split}
 \label{eqn:omgf}
 \end{equation}
The equation for $g_{j\sigma}$ is also written in the same way;
\begin{equation}
 \begin{split}
 \hat{\omega}g_{j\sigma} 
 &=
  \sum_{n}\epsilon^{(ba) (2,1)}_{jn}a_{n\sigma}+\sum_{m}\epsilon^{(bb) (2,1)}_{jm}b_{m\sigma}\\
 &\quad +\sum_{i}\epsilon^{(ba) (2,2)}_{ji}f_{i\sigma}+\sum_{m}\epsilon^{(bb) (2,2)}_{jm}g_{m\sigma}
 +R_{i\sigma}.
 \end{split}
 \end{equation}
The explicit forms of the remaining operators $L_{i\sigma}$ and $R_{j\sigma}$ will be given later.
The matrix $\epsilon^{(1,1)}$ is defined as
 \begin{equation}
 \epsilon^{(aa)(1,1)}_{in}
 = \tilde{U}\langle \eta_{-\sigma}^{a} \rangle \delta_{in}+ \lambda_{in}^{A},
    \label{eqn:e_aa}
 \end{equation}
 \begin{equation}   
 \epsilon^{(ab) (1,1)}_{ij}
 = -p_{ij}( \tilde{U} \langle \eta_{i-\sigma}^{a} \rangle 
              - \tilde{U} \langle \eta_{j-\sigma}^{b} \rangle),
 \end{equation}
 whereas $\epsilon^{(bb)(1,1)}$ is obtained from $\epsilon^{(aa)(1,1)}$ by replacing `a (A)' with `b(B)' and $p$ with $-p$. Here, $\epsilon^{(ba)(1,1)}$ is obtained from the symmetry property  \(\epsilon^{(ba)(1,1)}_{ji} = \epsilon^{(ab)(1,1)}_{ij}\).
\nopagebreak
The matrix $\epsilon^{(2,1)}$ is also derived up to the first order of $p$ as 
\begin{subequations}
\begin{equation}
 \epsilon^{(aa)(2,1)}_{in}
 = \langle (\delta \Phi)^{2}\rangle \delta_{in},
    \label{eqn:e2_aa}
 \end{equation}
 \begin{equation}   
 \epsilon^{(ab) (2,1)}_{ij}
 = -\langle p_{ij}( ( \delta \Phi_{i\sigma} )^{2} - ( \delta \Phi_{j\sigma} )^{2} ) \rangle.
 \end{equation}
 \label{eqn:e21A}
 \end{subequations}
 Here, however, the matrix $\epsilon^{(2,1)}$ need to have positive eignvalue since it is the norm of the operators $f_{i\sigma}$ and $g_{j\sigma}$. We modify it within the second order of $p$ to guarantee the positivity. It is modified, in the momentum representation, as  
 \begin{equation}   
 \epsilon^{(2,1)}_{k}
 =\dfrac{1}{1+(p_{k}^{2})}\left( \begin{array}{cc}
 1 & p_{k} \\
 -p_{k} & 1
 \end{array} \right)
 \left( \begin{array}{cc}
 \langle ( \delta \Phi_{i\sigma} )^{2} \rangle & 0 \\
 0 & \langle ( \delta \Phi_{j\sigma} )^{2} \rangle
 \end{array} \right)
 \left( \begin{array}{cc}
 1 & -p_{k} \\
 p_{k} & 1
 \end{array} \right),
 \label{eqn:e21B}
 \end{equation}
 where $p_{\mib k}$ is the Fourier transformation of $p_{ij}$.
 Since the right hand side of eq.(\ref{eqn:e21B}) is a unitary transformation of a diagonal matrix with positive matrix elements, $\epsilon^{(2,1)}$ in this form has non-negative eignvalue.  
 Note that eq.(\ref{eqn:e21B}) is equivalent to the previous eq.(\ref{eqn:e21A}) up to the first order of $p$.
 In order to write down $\epsilon^{(2,2)}$,             
 we define the matrix $m^{(2,2)}_{xy}$, which is related to the matrix $\epsilon^{(2,2)}$ with
 \begin{equation}
 \epsilon^{(2,2)}=m^{(2,2)}(\epsilon^{(2,1)})^{-1}.
 \end{equation}
 The matrix $m^{(2,2)}$ is given by
 \begin{align}
 m^{(aa)(2,2)}_{in}
 &= \left( 
 \tilde{U} 
    \left(1-\langle \eta_{-\sigma} \rangle \right)
 + \lambda_{00}^{A} \right) \epsilon^{(aa)(2,1)}_{in}\nonumber \\
  &\quad 
  -\delta \mu^{A}_{i\sigma}\delta_{in}+ \lambda_{in}^{A}{\cal C}^{A}_{in}
   \nonumber \\
 &\quad+\langle (\delta \Phi_{i\sigma})^{3} \rangle \delta_{in}.
 \label{eqn:aaa}
 \end{align}
 \begin{align}   
 m^{(ab) (2,2)}_{ij}
 &= -p_{ij} \langle ( {\cal T}_{i}-{\cal T}_{j} ) \delta \Phi_{j\sigma} \rangle
   -p_{ij} \langle {\cal T}_{i} \delta (\Phi_{i\sigma}-\Phi_{j\sigma}) \rangle
  \nonumber \\
 &\quad-\sum_{l}p_{il} \langle ( \delta \Phi_{i\sigma} - \delta \Phi_{l\sigma})\nonumber \\
                &\qquad \qquad (\lambda^{B}_{jl}+\Delta \Phi_{j\sigma} \delta_{jl}) 
                \delta \Phi_{l\sigma}) \rangle
 \nonumber \\
 &\quad-\sum_{n}p_{nj} \langle ( \delta \Phi_{n\sigma} - \delta \Phi_{j\sigma})
 \nonumber \\
 &\qquad \qquad
                (\Delta \Phi_{i\sigma}\delta_{in} +\lambda^{A}_{in})
                \delta \Phi_{i\sigma}  \rangle 
 \label{eqn:MabCDW}
 \end{align}
In these equations, we have introduced the following quantities;
 \begin{subequations}
 \begin{align}
 {\cal T}^{r}_{i\sigma} 
 &\equiv \tilde{U}{\cal J}^{r}_{i-\sigma}+\tilde{V}_{ij}({\cal J}^{\bar{r}}_{j\sigma} + {\cal J}^{\bar{r}}_{j-\sigma}),
 \\
 {\cal J}^{r}_{x\sigma}
 &\equiv
 \lambda^{r}_{xy}( d_{x\sigma}^{\dagger}d_{y\sigma} - d_{y\sigma}^{\dagger}d_{x\sigma}) \nonumber \\
 &\quad \mp p_{xy}(a_{x\sigma}^{\dagger}\lambda^{B}_{yz}b_{z\sigma}
                       -\lambda^{B}_{yz}b_{z\sigma}^{\dagger}a_{x\sigma}
                       \nonumber \\
 &\quad \qquad +b_{y\sigma}^{\dagger}\lambda^{A}_{xz}a_{z\sigma}
                       -\lambda^{A}_{xz}a_{z\sigma}^{\dagger}b_{y\sigma}
                       ), \\
  \delta \mu^{r} &\equiv \langle {\cal T}_{x\sigma}\delta \Phi_{x\sigma}(1-2n_{i\sigma})\rangle\nonumber \\
          &= \left( \tilde{U}^{2} {\cal K}_{x-\sigma} + 
                      \tilde{V}^{2}_{xy} ({\cal K}_{y\sigma} + {\cal K}_{y-\sigma}) \right) (1-2\langle n_{i\sigma}\rangle),\\
  {\cal K}^{r}_{x\sigma} &\equiv \sum_{y:y\neq x} \lambda^{r}_{xy} \langle d_{x\sigma}^{\dagger}d_{y\sigma} \rangle, \\
  {\cal C}^{r}_{xy\sigma} 
  &\equiv  
 \langle  \delta \Phi^{V}_{x} \delta \Phi^{V}_{y} \rangle
 + \tilde{U}^{2} 
 \langle \frac{1}{4}\delta \eta^{r}_{x}\delta \eta^{r}_{y}
   + \mbox{ \boldmath{$S$}$_{x}$ \boldmath{$S$}$_{y}$}
   -\Delta^{p \dagger}_{x}\Delta^{p}_{y} 
   \rangle.   
 \end{align}
 \end{subequations}
Here, the label `$r(\bar{r})$' indicates either `$A(B)$' or `$B(A)$', whereas $\mp$ takes $-$ sign if the index $x$ in the left hand side is in the `$A$' lattice, and $+$ if $x$ is in the `$B$' lattice. Greek indices $\mu,\nu$ represent both site and spin indices.
 From eq.~(65) $m^{(bb)}$ is also obtained by replacing $a(A)$ with $b(B)$ and $p$ with $-p$. 
 The matrix $m^{(ba)}$ is obtained by the symmetry relationship $m^{(ba)}_{ji} = m^{(ab)}_{ij}$. 
 By using these quantities, the remaining operator $L_{i\sigma}$ in eq.(\ref{eqn:omgf}) is expressed as
 \begin{equation}
L_{i\sigma}={\cal T}_{i\sigma}a_{i\sigma}
            +\delta \Phi_{i\sigma} \sum_{n}\lambda^{A}_{in}a_{n\sigma}
            -\sum_{j}p_{ij}({\cal T}_{i}-{\cal T}_{j})b_{j\sigma}
            +L^{\prime}_{i\sigma}
            -\sum_{n}\tilde{\lambda}^{A}_{in}f_{n\sigma}, 
\end{equation}
where $\tilde {\lambda}^{A}$ is defined as \( \tilde{\lambda}^{A}_{in} \equiv \sum_{l}\left( -\delta \mu^{A}\delta_{il}+{\cal C}^{A}_{il}\lambda^{A}_{il}\right)
\left( (\epsilon^{(a,a)(2,1)})^{-1}\right)_{ln}. \)
Here, $L^{\prime}_{i\sigma}$ comes from higher order fluctuation terms as
\begin{equation}
 L^{\prime}_{i\sigma}= \delta {\cal Q}_{i\sigma} a_{i\sigma} -\sum_{j} p_{ij}( \delta {\cal Q}_{i\sigma}- \delta {\cal Q}_{j\sigma})b_{j\sigma}.
 \end{equation}
Here, $\delta {\cal Q}$ is defined as \(\delta {\cal Q}_{i\sigma}\equiv (\delta \Phi_{i\sigma})^{2} - \dfrac{\langle (\delta \Phi_{i\sigma})^{3} \rangle}{\langle (\delta \Phi_{i\sigma})^{2}\rangle} \delta \Phi_{i\sigma} \) as the same as in the previous subsection.
 We ignore them in the same manner as in the previous subsection. The operator $R_{j\sigma}$ is also written in the same way as
\begin{equation}
R_{j\sigma}={\cal T}_{j\sigma}b_{j\sigma}
            +\delta \Phi_{j\sigma} \sum_{m}\lambda^{B}_{jm}b_{m\sigma}
            -\sum_{i}p_{ij}({\cal T}_{i}-{\cal T}_{j})a_{i\sigma}
            +R^{\prime}_{j\sigma}
            -\sum_{m}\tilde{\lambda}^{B}_{jm}g_{m\sigma}. 
\end{equation}
Here, $R^{\prime}_{j\sigma}$ is given as
\begin{equation}
 R^{\prime}_{j\sigma}= \delta {\cal Q}_{j\sigma} b_{j\sigma} -\sum_{i}p_{ji}( \delta {\cal Q}_{i\sigma}- \delta {\cal Q}_{j\sigma})a_{i\sigma},
 \end{equation}
and it is neglected here.

 This formulation is valid only if the CDW order parameter is large enough to ignore the higher order of the expansion parameter $p\equiv\frac{t}{2\Delta_{\scriptscriptstyle CDW}}$. However, we expect that this formulation gives better result even when the order parameter amplitude is modest since the CDW basis becomes correct in the strong coupling limit up to the second order in $\frac{t}{V}$. 
 If the intersite interaction is large so that $t \ll \Delta_{CDW}$, one can expand the quasiparticle dispersion eq.(\ref{eqn:eignvalue}) up to the second order in $\frac{t}{\Delta_{CDW}}$ as 
 \[\lambda_{\mib k}=\Delta_{\scriptscriptstyle CDW}+\frac{t_{\mib k}^{2}}{2\Delta_{\scriptscriptstyle CDW}}.\]
 Performing Fourier transformation, one can obtain the effective hopping as \[\lambda_{i,j}=\frac{t^{2}}{2\Delta_{CDW}}\beta_{i,j}\] where \(\beta_{i,j} \equiv \frac{1}{N}\sum_{k} e^{i \mib{k}(\mib{r}_{i}-\mib{r}_{j})}(2(\cos(k_x)+\cos(k_{y}))^{2} \) is the next-nearest neighbor hopping. This corresponds to the second order perturbation around the charge-ordered state.
 We refer to this formulation as `CDW basis approach' in contrast to the formulation introduced in the previous subsection `standard approach', which takes $\{c_{i\sigma}\}$ as the operator basis.


\section{Numerical Results}
In this section, we show the result of the numerical calculations.

 The following calculations have been performed on the $32\times32$ lattice with Matsubara frequencies 2048. 
 The spectral functions have been obtained by analitic continuation using Pad\'{e} approximation~\cite{Pade}. The obtained real frequency data are smoothened by taking convolution with the Gaussian distribution function \( N(\omega ; \sigma)\equiv \frac{1}{ \sqrt{2\pi}\sigma }e^{-\frac{\omega^{2}}{2\sigma^{2}}} \) with $\sigma=0.05$.
 
 Through this paper the energy unit is set as $t=1$. 
First, we show the phase diagram of the extended Hubbard model obtained by CPM. We first derived the charge-ordering transition line by the `standard approach', taking ${c_{i}}$ as the operator basis. We then have changed the basis to the `CDW basis' in order to further analyze the region where the CDW order parameter $\Delta_{\rm CDW}$ is large.

First, we compare the phase diagram obtained by CPM with the corresponding one
 obtained by the Hartree-Fock approximation (HFA). Here, the both calculations were performed allowing $2\times2$ sublattices. In the HFA calculation, we have examined longer-ranged ordering patterns allowing up to $4\times4$ sublattices but no other type of order was obtained as a stable phase.
 One clear difference of the present CPM result from the corresponding result by HFA is seen in the existence of the insulating phase with charge order but without the magnetic order (non-magnetic insulating phase). 
 Here we note that, even in the CPM calculation, antiferromagnetic order is indeed obtained at low temperatures. 
The obtained ordering pattern is shown in Fig.~1, which is in agreement with strong-coupling analysis as well as the phase $(3)$ of the HFA phase diagram.
 At first thought, by taking account the fact we are considering two-dimensional system, we can argue that this result violates Mermin-Wagner theorem~\cite{Mermin}. This violation is, however, attributed to the mean-field like nature of the calculation method. We can expect that the problem would be reduced if we take larger sublattices.

Here, we emphasize that in the CPM calculation, magnetic order is suppressed to low temperatures. In Fig~\ref{fig:TNplot} we have shown the N\'{e}el temperature for several parameters.  This low ordering temperatures indicate that the magnetic order is induced by a small energy scale $J_{\rm eff}$, which is the exchange energy in the charge-ordered phase. The explicit calculation of this energy will be given in the Appendix~B. Here, the calculated $T_N$ is almost the same order as the exchange energy scale obtained by the strong coupling expansion. Although $T_N$ is higher than $J_{\rm eff}$ especially in the weak coupling region, this is atttributed to the deviation from strong coupling assumption.
\begin{figure}
\setlength{\unitlength}{1.2mm}
\begin{center}
\begin{picture}(20,20)
\put(7.5,-2.5) {\makebox(0,0)[cc]{$T<T_{N}$}}
\put(2.5,0.0) {\thicklines \vector(0,1){5}}
\put(12.5, 0.0) {\thicklines \vector(0,1){5}}
\put(7.5, 10.0) {\thicklines \vector(0,-1){5}}
\put(2.5,10) {\thicklines \vector(0,1){5}}
\put(12.5,10) {\thicklines \vector(0,1){5}}
\put(0.0,2.5) {\line(1,0){15}}
\put(0.0,7.5) {\line(1,0){15}}
\put(0.0,12.5) {\line(1,0){15}}
\put(2.5,0.0) {\line(0,1){15}}
\put(7.5,0.0) {\line(0,1){15}}
\put(12.5,0.0) {\line(0,1){15}}
\end{picture}
\begin{picture}(20,20)
\put(7.5,-2.5) {\makebox(0,0)[cc]{$T>T_{N}$}}
\put(2.5,2.5) {\circle{2}}
\put(12.5, 2.5) {\circle{2}}
\put(2.5,12.5) {\circle{2}}
\put(12.5,12.5) {\circle{2}}
\put(7.5,7.5) {\circle{2}}
\put(0.0,2.5) {\line(1,0){15}}
\put(0.0,7.5) {\line(1,0){15}}
\put(0.0,12.5) {\line(1,0){15}}
\put(2.5,0.0) {\line(0,1){15}}
\put(7.5,0.0) {\line(0,1){15}}
\put(12.5,0.0) {\line(0,1){15}}
\end{picture}
\end{center}
\label{fig:NeelORDER}
\caption{
Charge and/or spin distribution of the ordered patterns obtained by CPM (with CDW basis).
Large circles indicate charge rich sites, while arrows indicate spin polarized sites. Left panel shows the solution for $T<T_N$ with magnetic order. This pattern is the same as the HFA solution~(3). Right panel shows the solution for $T>T_N$ without magnetic order.  
}
\end{figure}
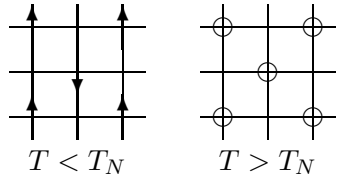

On the other hand, metal-insulator transition as well as charge-ordering transition occur in larger energy scale. We have confirmed that the charge order as well as the metal-insulator transition up to $T \approx 0.5$, which is several factor higher than $T_N$.
This is in contrast to HFA, in which magnetic long-range order is necessary to reduce double occupancy. Thus magnetic order is induced by larger energy scale $U$, and it exists even in relatively high temperature. 

Using CPM (CDW basis approach) and taking account of short-range correlation effects, we can obtain a phase with small double occupancy but without magnetic long-range order. This correspond to the non-magnetic insulating phase.
\begin{figure}
\begin{center}
\includegraphics[trim=0mm 0mm 0mm 0mm, clip, scale=0.75, angle=-90]{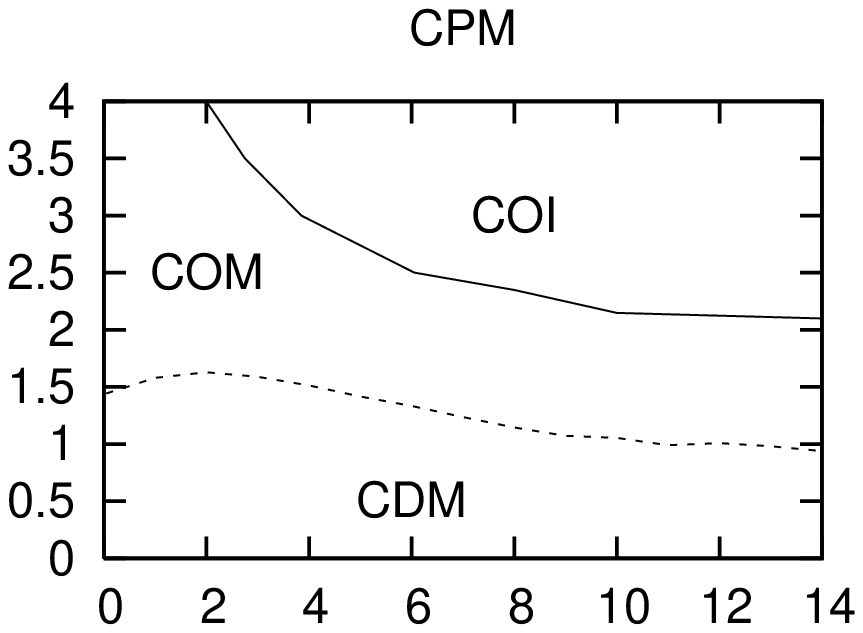}
\includegraphics[trim=0mm 0mm 0mm 0mm, clip, scale=0.75, angle=-90]{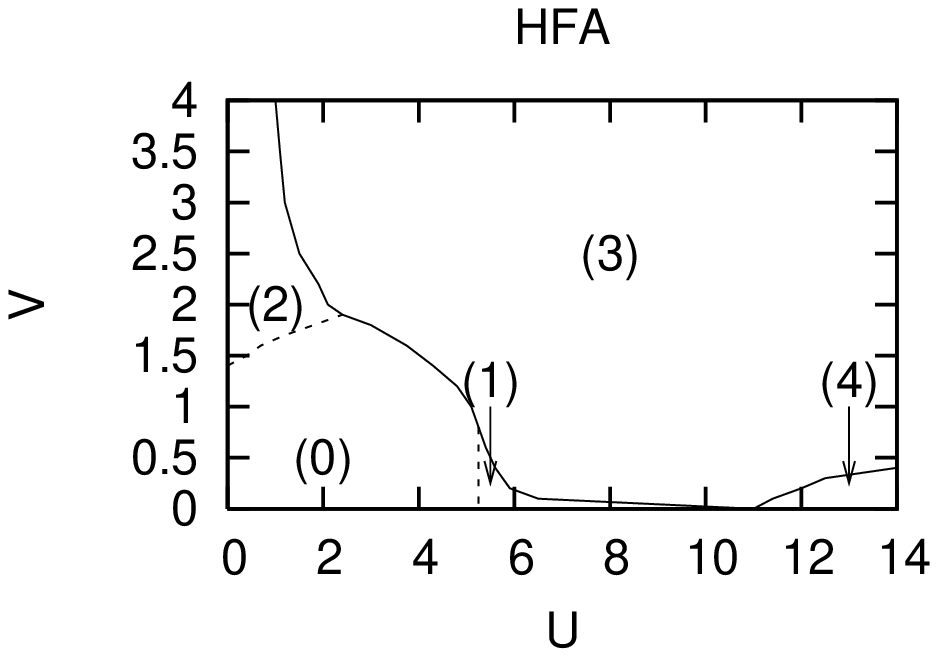}
\caption{\
Phase diagram for the extended Hubbard model at quarter filling. The left panel shows the result obtained by CPM, while the right panel shows the corresponding result obtained by HFA. Temperature is set $T=0.15$. The dotted lines indicate continuous transitions, while the solid lines indicate discontinuous transitions. In the left panel, COI, COM, and CDM indicate the charge-ordered insulating, the charge-ordered metallic, and the charge-disordered metallic phases, respectively. In the right panel, the labels (0) to (4) indicate different ordering patterns of spin and/or charge obtained by the HFA calculation. Each ordering pattern is shown in Appendix~B.  
}
\end{center}
\label{fig:CPMphase}
\end{figure}
\begin{figure}
\begin{center}
\includegraphics[trim=0mm 0mm 0mm 0mm, clip, scale=0.75, angle=-90]{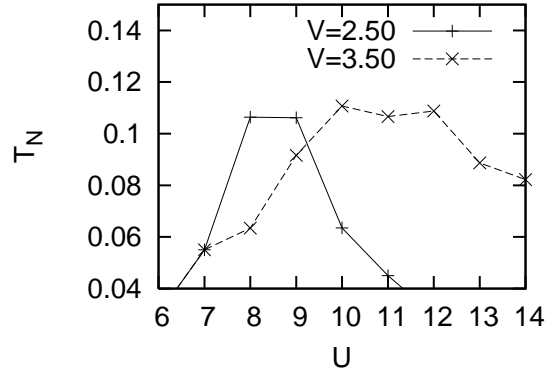}
\caption{\
Plot of the N\'{e}el temperature $T_N$ as a function of interaction parameter $U$, $V$.  The result is obtained by CPM based on CDW basis.
}
\label{fig:TNplot}
\end{center}
\end{figure}
 Using HFA, the square lattice system at quarter filling become insulating only if there is a spin symmetry breaking which splits the band into four. On the other hand, by taking account of dynamical correlation effects as in CPM, one can reproduce insulating phase without magnetic symmetry breaking.

 We note that application of CPM in the `standard approach' does not reproduce the metal-insulator transition. This is because the operator projection analysis up to the second-order expansion cannot fully reproduce the correct dynamics around the charge-ordering transition. 
On the other hand, the formulation based on the CDW basis reproduces the metal-insulator transition. This is attributed to the fact that this formulation recovers the strong coupling limit up to the second order in $\frac{t}{V}$ as we have seen before. This basis function gives qualitatively correct results even at the truncation up to the second order.
 
 Here, one should note that the two approaches do not contradict at intermediate coupling since they are connected with (approximate) unitary transformation.
 Therefore, one can interpolate these two approaches and can obtain qualitatively correct results in the whole parameter space. In Fig.~4, we have plotted the spectral function obtained by the both approaches. The two spectra show similar structure for intermediate amplitude of $V$.
\begin{figure}[tdp]
\begin{center}
\includegraphics[trim=0mm 0mm 0mm 0mm, clip, scale=0.75, angle=-90]{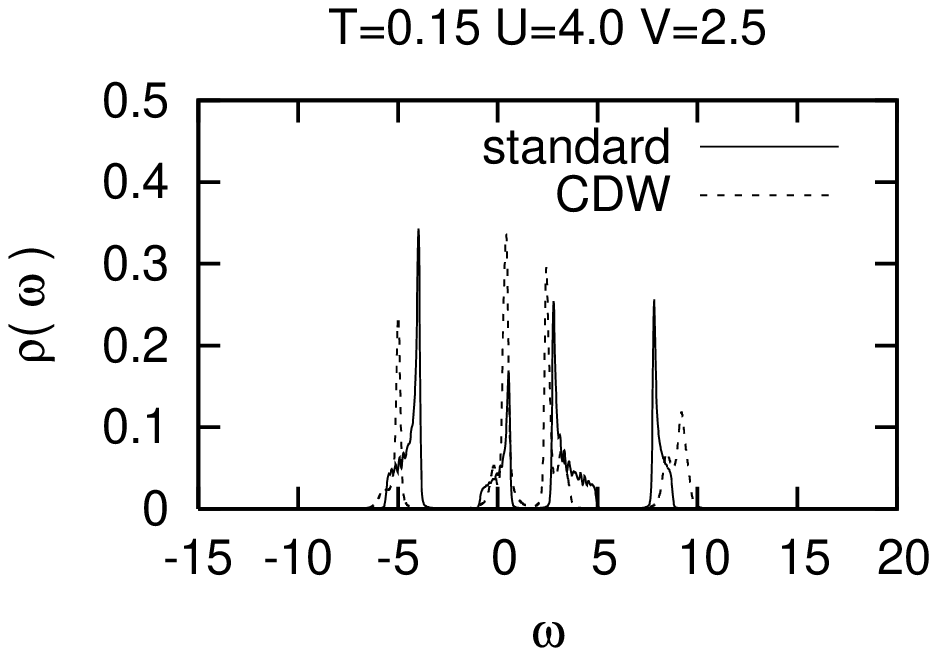}
\includegraphics[trim=0mm 0mm 0mm 0mm, clip, scale=0.75, angle=-90]{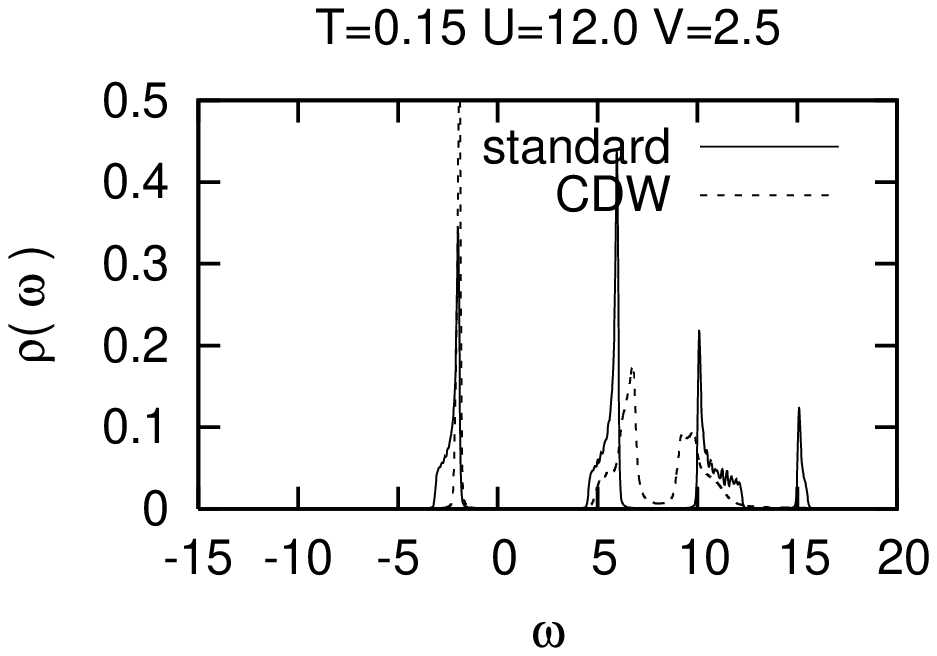}
\includegraphics[trim=0mm 0mm 0mm 0mm, clip, scale=0.75, angle=-90]{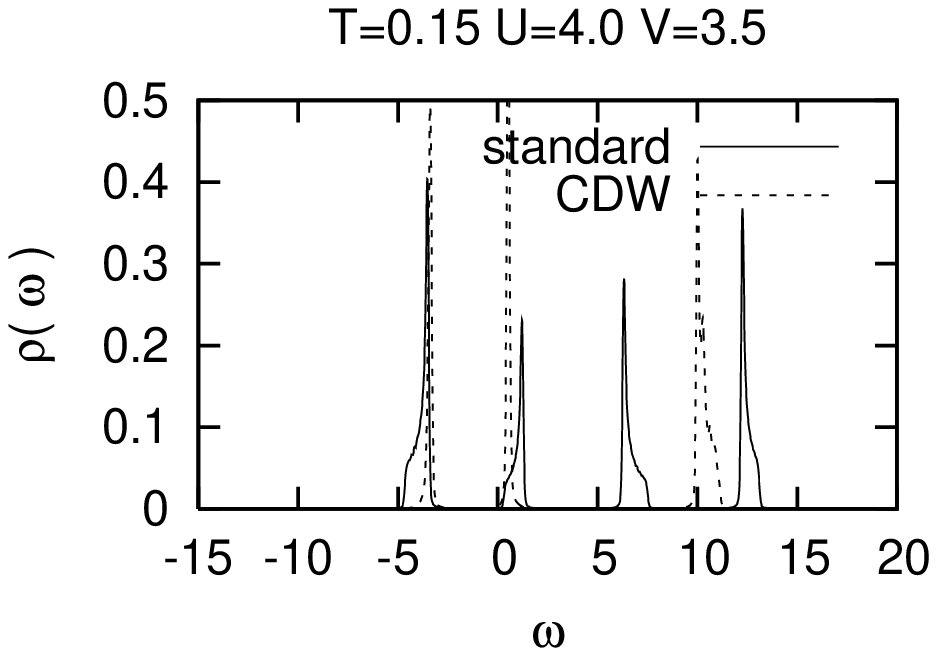}
\includegraphics[trim=0mm 0mm 0mm 0mm, clip, scale=0.75, angle=-90]{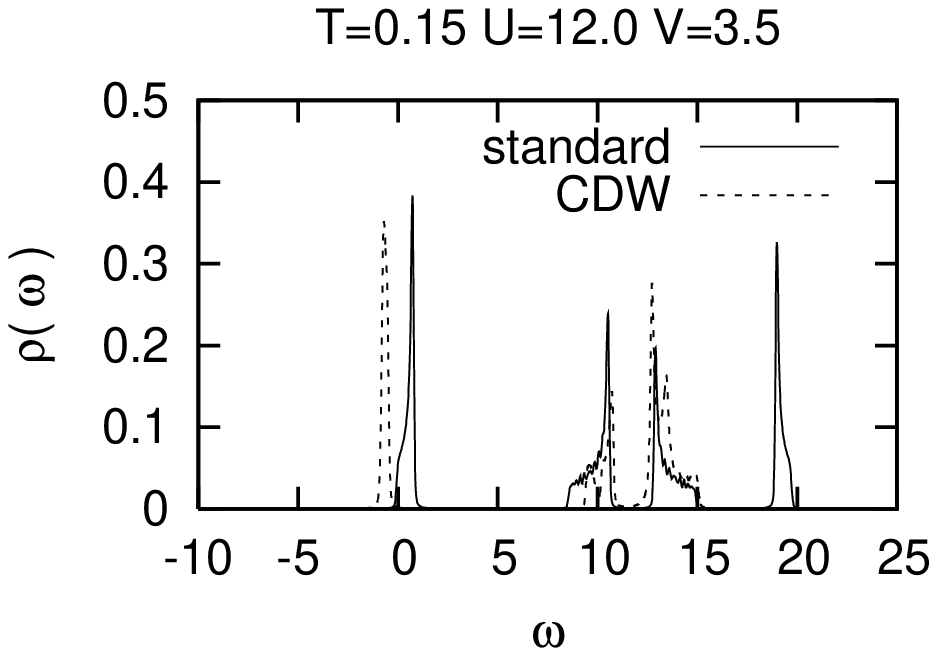}
\caption{Local density of states $\rho(\omega)$ of the extended Hubbard model at quarter filling. Temperature is set $T=0.15$, while the interaction parameters are set $V=2.5$, $U=4.0,12.0$, and $V=3.5$,$U=4.0,12.0$ respectively. Solid line indicates the result obtained by the CPM based on standard basis, while the dotted line indicates the result obtained by the CPM based on CDW basis.}
\end{center}
\label{fig:compspec}
\end{figure}

\begin{figure}
\includegraphics[trim=0mm 0mm 0mm 0mm, clip, scale=0.7, angle=-90]{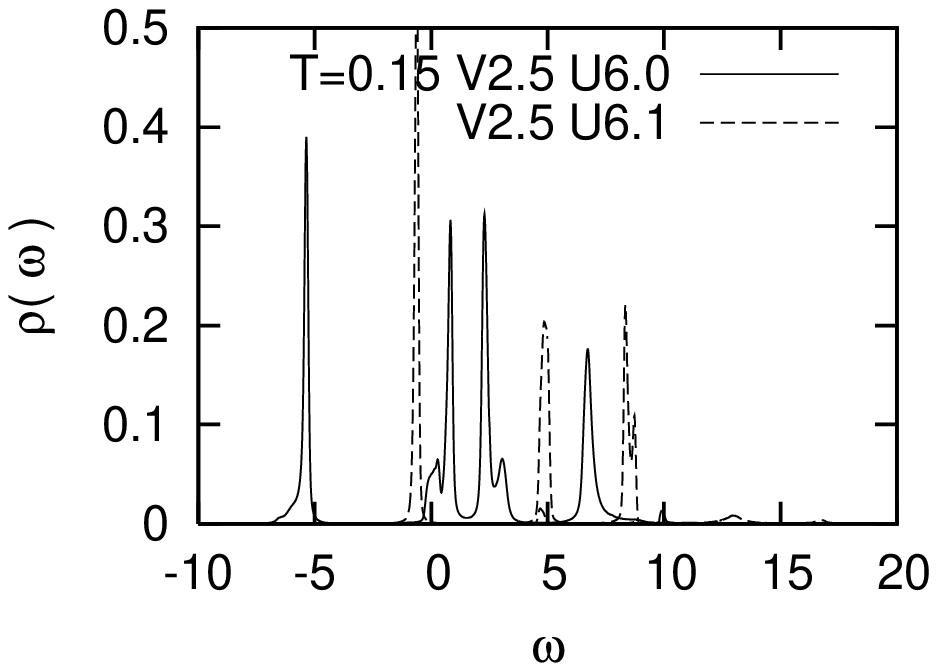}
\includegraphics[trim=0mm 0mm 0mm 0mm, clip, scale=0.7, angle=-90]{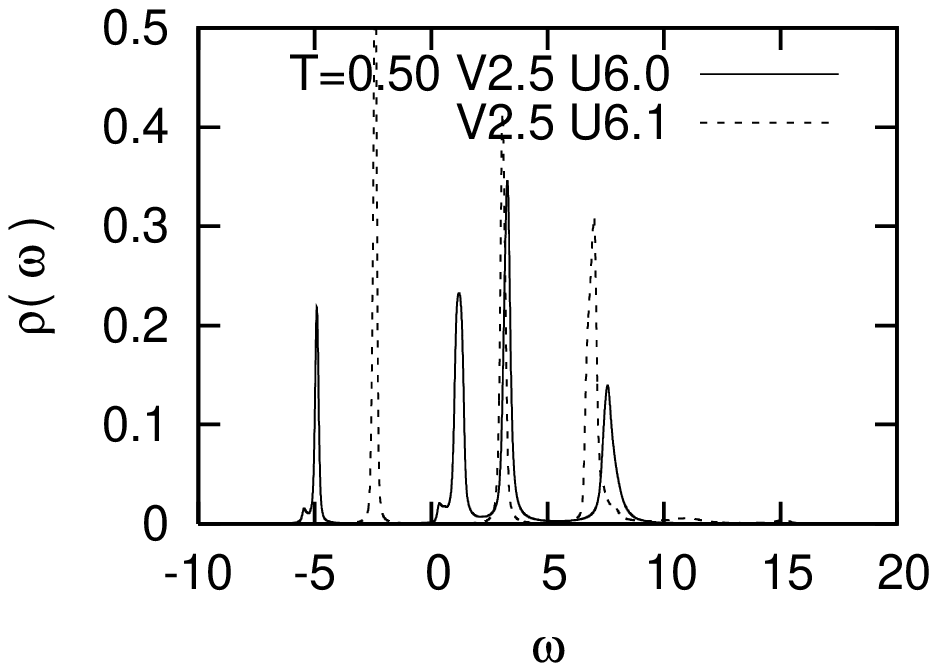}
\caption{Local density of states $\rho(\omega)$ at the metal-insulator transition point, The result is obtained by CPM based on CDW basis. Left panel shows the result for $T=0.15$. The parameter is set as $V=2.5$ $T=0.15$, and $U=6.0$, $U=6.1$ respectively. The solid line indicates the metallic solution while the dotted line indicates the insulating solution. The right panel shows the result for $T=0.50$.}
\label{fig:MITspec}
\end{figure}

 In Fig.~\ref{fig:MITspec}, we show the spectral function of the metallic as well as the insulating phase. The former phase has a large density of states at the Fermi level, while the latter phase has small density of states, which becomes vanishing in the zero temperature limit.
 
\begin{figure}
\includegraphics[trim=0mm 0mm 0mm 0mm, clip, scale=0.7, angle=-90]{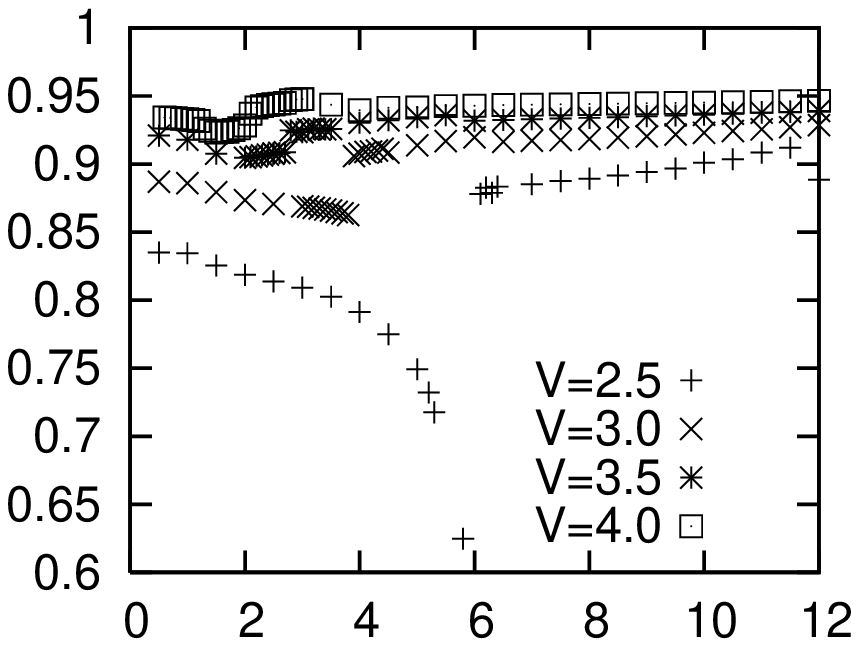}
\includegraphics[trim=0mm 0mm 0mm 0mm, clip, scale=0.7, angle=-90]{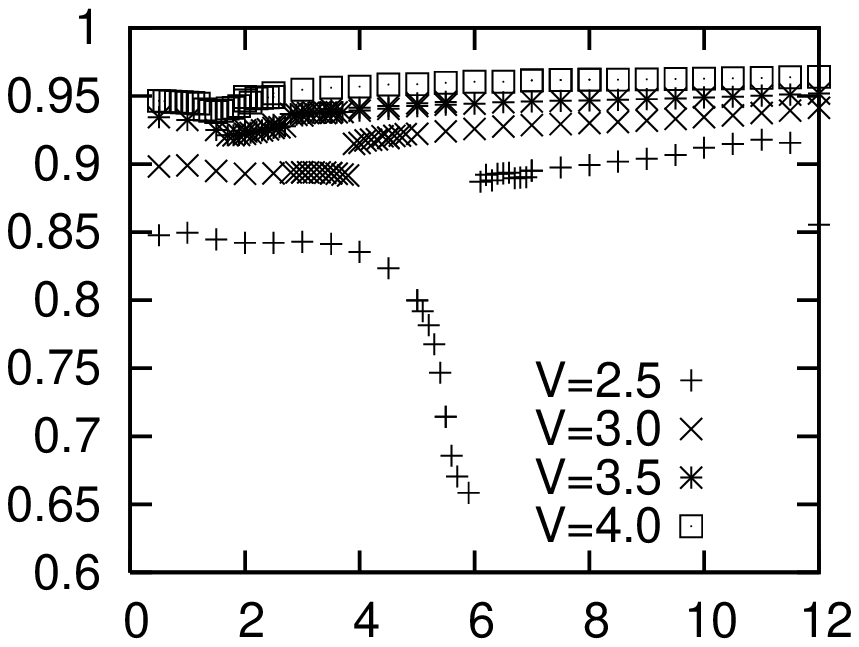}
\caption{Order parameter of the charge-ordered phase $\langle \Delta n \rangle$ plotted as a function of on-site interaction $U$.Temperature is set $T=0.15$ (left panel) and $T=0.50$ (right panel), respectively.}
\label{fig:orderjump}
\end{figure}

 Our result shows that the transition is of the first order, which is typical to (finite temperature) metal-insulator transitions. The order parameter of the CDW state $\langle \Delta n\rangle \equiv \langle n_{A} \rangle -\langle n_{B} \rangle $ jumps at the transition point as is shown Fig.~\ref{fig:orderjump}

At fixed $V$, the order parameter $\langle \Delta n \rangle$ decreases as a function of $U$ in the metallic phase, but it takes nearly a constant value in the insulating phase. This is explained as follows; in the metallic phase with delocalized wave function, charge disproportionation increase the double occupancy and thus is disfavored in terms of the onsite repulsion term. On the other hand, in the insulating phase where the charges are localized each per one site, charge disproportionation does not severely increase the double occupancy. 
Moreover, suppression of the charge fluctuations results in larger charge disproportionation in the insulating phase.

In this picture, however, the insulating phase occur only in the charge-ordered phase. This means that there exist charge-ordered metallic phase for a certain parameter region. On the other hand, such phases are not widely observed in experiments. This issue will be discussed in the next section.

\section{Comparison to experimental observations}
In this section, we compare the obtained results with the experimental observations.

As we have mentioned in Sec.1, charge ordering occurs in various types of strongly correlated electron systems. Among them, charge ordering induced by the intersite Coulomb interactions is considered to be realized in a certain class of organic conductors~\cite{Seo,HubbardEHM}. Here, we note that there are number of transition metal oxides (TMO) that shows charge ordering. There are, however, some difficulties in comparing those materials with this result since in many cases, TMO's have more than one band and the spin degree of freedom is strongly coupled to the orbital degree of freedom. The result obtained here, by simple one band model with only electron-electron repulsion, may not be applicable. 
On the other hand, many of organic conductors have single conduction band and the result of single-band model is directly applicable. In addition to this, these materials provide good examples where the charge ordering temperature is much higher than the temperature at which spin degree of freedom is lost (magnetic-ordering temperature, spin Peierls temperature etc.) as we will see below.

  For example, $\theta$-(BEDT-TTF)$_{2}$MM$^{\prime}$(SCN)$_{4} $(M=Rb,Cs, M$^{\prime}$=Zn,Co) have large two-dimensional anisotropy and strong electron-electron correlations compared to the bandwidth~\cite{Seo,HubbardEHM,Moritake1,Moritake2}. In addition to this, they have weak dimerization and regarded as quarter (hole-) filling system. Thus they offer good candidates of the charge-ordering system. In fact, $\theta-$(BEDT-TTF)$_{2}$RbZn(SCN)$_{4}$, for example, undergoes MIT at $T_{\rm MIT}=195K$ at ambient pressure, where the electrical resistivity shows sharp increase, with no sharp change in the magnetic susceptibility until a magnetic transition to spin Peierls state at $T_{\rm SP}=50K$ ~\cite{Mori98}. The evidence of charge disproportionation was reported from the data of $ ^{13}$C NMR shift and relaxation time measurement~\cite{Miyagawa00}. 
  Between $195K$ and $50K$, an insulating phase without any magnetic order is observed, which is in agreement with our numerical calculation.
  The dynamical as well as spatial correlations are considered to be important in the above-mentioned system.The remaining magnetic degrees of freedom is lost through the spin-Peierls transition in the low temperature phase. This transition is considered to be induced by the coupling to the lattice degrees of freedom.
  Here, however, we do not further discuss this problem since we do not take account of lattice degrees of freedom. In this paper, we have adopted purely electronic model in order to study the physics in charge degrees of freedom. Study of the electron-lattice interplay remains as a future work.\\

  Here, it should be noted that the ordering pattern observed in $\theta-$(BEDT-TTF)$_{2}$RbZn(SCN)$_{4}$ is not the checkerboard order but of the stripe order. This difference, however, may be attributed to the difference of the lattice shape. In the real system, there exist finite values of the Coulomb interactions as well as the hopping in the diagonal directions. 
  
  More important difference is the existence of the charge-ordered metallic phase in the numerical calculations, while it is not widely observed in experiments. This issue has been pointed out in Sec.1.
  One possible reason for the difference is the overestimate of the charge-ordered phase in the numerical calculation. Although CPM takes into account spatial short-ranged correlation effects, it still has a mean-field like character. This may result in the underestimate of spatial fluctuations and overestimate of charge order.
  If one could treat the spatial fluctuations more accurately, the phase boundary between the charge-ordered phase and charge-disordered phase may merge to the phase boundary between the metallic and insulating phase. 
On the other hand, however, this problem might also be attributed to the limitation of the extended Hubbard model itself as a realistic model of real materials. In real materials, charge ordering often accompanies some additional symmetry breaking such as lattice distortion or dimerization, which may further stabilize the insulating phase. 
  
\section{Summary}
We have investigated the two-dimensional extended Hubbard model with the correlator projection method (CPM). 
CPM is a newly developed method which can treat correlation effects in a self-consistent manner.

Applying CPM, we have shown that a metal-insulator transition is induced by charge ordering. The insulating phase has antiferromagnetic order below $T_{N}$, which is as the same order as the exchange interaction energy $J_{\rm eff}$, but is non-magnetic insulator above $T_{N}$. This is consistent with the strong-coupling picture as well as several experimental observations. The temperature scale of $T_{N}$ is several factor smaller than the charge-ordering temperature $T_{CO}$ and/or metal-insulator transition temperature $T_{MIT}$.
We also emphasize that the obtained result is in contrast to the Hartree-Fock approximation. The difference of energy scale as well as the existence of non-magnetic insulating phase is obtained by taking account of the dynamical and spatial correlation effects. 

However, we have also obtained a charge-ordered metallic phase that is not widely observed in real materials. 
The difference may arise from the insufficient treatment of spatial fluctuation effects. More accurate estimate of short-ranged spatial fluctuation effects is left for future studies.
The emergence of charge-ordered metal in the present result in general contradiction to the experimental observations may also be attributed to the limitation of the extended Hubbard model itself, where the coupling to lattice is ignored.
 
Although we clearly need further improvement, we have succeeded in obtaining a new formulation that can treat metal-insulator transitions induced by the charge ordering.
The result also indicates that the qualitative aspect of the experimentally-observed metal-insulator transitions in charge-ordering systems can be reproduced by a simple model as the extended Hubbard model if one carefully treat the correlation effects.

\section{Acknowledgements}
One of the authors ( K.H.) would like to thank S. Onoda, S. Watanabe, and Y. Imai for stimulating discussions.

\appendix
\section{Phases obtained by the Hartree-Fock approximation}
The pattern of the charge and/or spin obtained by the Hartree-Fock approximation is shown below. We have examined possible spatial pattern up to $4\times 4$ period and obtained the following 5 phases as stable ones. We do not exclude the possibility of obtaining lower energy state by allowing spatial patterns with longer period than $4\times 4$.
\nopagebreak
\begin{figure}[hb]
\setlength{\unitlength}{1.2mm}
\begin{picture}(20,20)
\put(7.5,-2.5) {\makebox(0,0)[cc]{\small(0)}}
\put(2.5,2.5) {\circle{1}}
\put(12.5, 2.5) {\circle{1}}
\put(2.5,12.5) {\circle{1}}
\put(12.5,12.5) {\circle{1}}
\put(7.5,7.5) {\circle{1}}
\put(2.5,7.5) {\circle{1}}
\put(12.5, 7.5) {\circle{1}}
\put(7.5,2.5) {\circle{1}}
\put(7.5,12.5) {\circle{1}}
\put(0.0,2.5) {\line(1,0){15}}
\put(0.0,7.5) {\line(1,0){15}}
\put(0.0,12.5) {\line(1,0){15}}
\put(2.5,0.0) {\line(0,1){15}}
\put(7.5,0.0) {\line(0,1){15}}
\put(12.5,0.0) {\line(0,1){15}}
\end{picture}
\begin{picture}(20,20)
\put(7.5,-2.5) {\makebox(0,0)[cc]{\small(1)}}
\put(2.5,0.0) {\thicklines \vector(0,1){5}}
\put(2.5,5.0) {\thicklines \vector(0,1){5}}
\put(12.5,0.0) {\thicklines \vector(0,1){5}}
\put(12.5,5.0) {\thicklines \vector(0,1){5}}
\put(7.5, 5.0) {\thicklines \vector(0,-1){5}}
\put(7.5, 10.0) {\thicklines \vector(0,-1){5}}
\put(7.5, 15.0) {\thicklines \vector(0,-1){5}}
\put(2.5,10) {\thicklines \vector(0,1){5}}
\put(12.5,10) {\thicklines \vector(0,1){5}}
\put(0.0,2.5) {\line(1,0){15}}
\put(0.0,7.5) {\line(1,0){15}}
\put(0.0,12.5) {\line(1,0){15}}
\put(2.5,0.0) {\line(0,1){15}}
\put(7.5,0.0) {\line(0,1){15}}
\put(12.5,0.0) {\line(0,1){15}}
\end{picture}
\begin{picture}(20,20)
\put(7.5,-2.5) {\makebox(0,0)[cc]{\small(2)}}
\put(2.5,2.5) {\circle{2}}
\put(12.5, 2.5) {\circle{2}}
\put(2.5,12.5) {\circle{2}}
\put(12.5,12.5) {\circle{2}}
\put(7.5,7.5) {\circle{2}}
\put(0.0,2.5) {\line(1,0){15}}
\put(0.0,7.5) {\line(1,0){15}}
\put(0.0,12.5) {\line(1,0){15}}
\put(2.5,0.0) {\line(0,1){15}}
\put(7.5,0.0) {\line(0,1){15}}
\put(12.5,0.0) {\line(0,1){15}}
\end{picture}
\setlength{\unitlength}{1.2mm}
\begin{picture}(20,20)
\put(7.5,-2.5) {\makebox(0,0)[cc]{\small(3)}}
\put(2.5,0.0) {\thicklines \vector(0,1){5}}
\put(12.5, 0.0) {\thicklines \vector(0,1){5}}
\put(7.5, 10.0) {\thicklines \vector(0,-1){5}}
\put(2.5,10) {\thicklines \vector(0,1){5}}
\put(12.5,10) {\thicklines \vector(0,1){5}}
\put(0.0,2.5) {\line(1,0){15}}
\put(0.0,7.5) {\line(1,0){15}}
\put(0.0,12.5) {\line(1,0){15}}
\put(2.5,0.0) {\line(0,1){15}}
\put(7.5,0.0) {\line(0,1){15}}
\put(12.5,0.0) {\line(0,1){15}}
\end{picture}
\begin{picture}(20,20)
\put(7.5,-2.5) {\makebox(0,0)[cc]{\small(4)}}
\put(2.5,0.0) {\thicklines \vector(0,1){5}}
\put(2.5,5.0) {\thicklines \vector(0,1){5}}
\put(12.5,0.0) {\thicklines \vector(0,1){5}}
\put(12.5,5.0) {\thicklines \vector(0,1){5}}
\put(7.5, 0.0) {\thicklines \vector(0,1){5}}
\put(7.5, 5.0) {\thicklines \vector(0,1){5}}
\put(2.5, 10.0) {\thicklines \vector(0,1){5}}
\put(7.5, 10.0) {\thicklines \vector(0,1){5}}
\put(12.5, 10.0) {\thicklines \vector(0,1){5}}
\put(0.0,2.5) {\line(1,0){15}}
\put(0.0,7.5) {\line(1,0){15}}
\put(0.0,12.5) {\line(1,0){15}}
\put(2.5,0.0) {\line(0,1){15}}
\put(7.5,0.0) {\line(0,1){15}}
\put(12.5,0.0) {\line(0,1){15}}
\end{picture}
\caption{
Charge and/or spin distribution of the ordered patterns obtained by HFA.
Large circles indicate charge rich sites, while arrows indicate spin polarized sites. (0): no order
(1):$(\pi,0)$ spin order (2):$(\pi,\pi)$ charge order
(3):$(\pi,\pi)$ charge order with $(\pi,0)$ $(0,\pi)$ spin order 
(4):ferromagnetic order
}
\label{fig:HFpattern}
\end{figure}
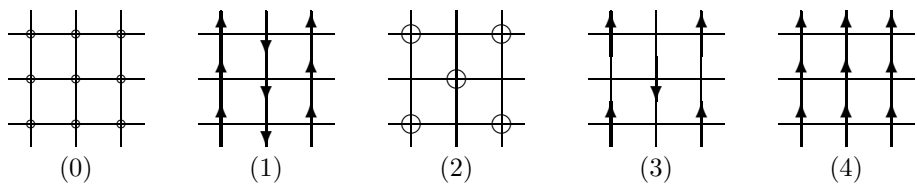

\section{Derivation of exchange coupling parameter}
\begin{figure}
\setlength{\unitlength}{2mm}
\begin{picture}(20,20)
\put(2.5,2.5) {\circle{2}}
\put(12.5, 2.5) {\circle{2}}
\put(2.5,12.5) {\circle{2}}
\put(12.5,12.5) {\circle{2}}
\put(7.5,7.5) {\circle{2}}
\put(0.0,2.5) {\line(1,0){15}}
\put(0.0,7.5) {\line(1,0){15}}
\put(0.0,12.5) {\line(1,0){15}}
\put(2.5,0.0) {\line(0,1){15}}
\put(7.5,0.0) {\line(0,1){15}}
\put(12.5,0.0) {\line(0,1){15}}
\put(6.0,-2.5) {\mbox{(a)}}
\end{picture}
\begin{picture}(20,20)
\put(2.5,2.5) {\circle{2}}
\put(12.5, 2.5) {\circle{2}}
\put(2.5,12.5) {\circle{2}}
\put(12.5,12.5) {\circle{2}}
\put(12.5,7.5) {\circle{2}}
\put(10.0,7.8) {\oval(4,4)[t]}
\put(12.0,7.8) {\vector(0,-1){0.3}}
\put(0.0,2.5) {\line(1,0){15}}
\put(0.0,7.5) {\line(1,0){15}}
\put(0.0,12.5) {\line(1,0){15}}
\put(2.5,0.0) {\line(0,1){15}}
\put(7.5,0.0) {\line(0,1){15}}
\put(12.5,0.0) {\line(0,1){15}}
\put(6.0,-2.50) {\mbox{(b)}}
\end{picture}
\begin{picture}(40,20)
\put(0.0,5.0) {\line(1,0){9.5}}
\put(0.0,10.0) {\line(1,0){9.5}}
\put(2.5,2.5) {\line(0,1){10}}
\put(7.5,2.5) {\line(0,1){10}}
\put(2.5,7.5) {\thicklines \vector(0,1){5.0}}
\put(7.5,7.5) {\thicklines \vector(0,-1){5.0}}
\put(4.0,0.0) {\mbox{(c)}}
%
\put(10.5,5.0) {\line(1,0){9.5}}
\put(10.5,10.0) {\line(1,0){9.5}}
\put(12.5,2.5) {\line(0,1){10}}
\put(17.5,2.5) {\line(0,1){10}}
\put(12.0,2.5) {\thicklines \vector(0,1){5.0}}
\put(13.0,7.5) {\thicklines \vector(0,-1){5.0}}
\put(11.8,7.5) {\oval(4,4)[l]}
\put(11.8,5.5) {\vector(1,0){0.3}}
\put(15.0,4.7) {\oval(3.6,3.6)[b]}
\put(13.2,4.7) {\vector(0,1){0.3}}
\put(14.0,0.0) {\mbox{(c$^{\prime}$1)}}
%
\put(20.5,5.0) {\line(1,0){9.5}}
\put(20.5,10.0) {\line(1,0){9.5}}
\put(22.5,2.5) {\line(0,1){10}}
\put(27.5,2.5) {\line(0,1){10}}
\put(27.0,2.5) {\thicklines \vector(0,1){5.0}}
\put(28.0,7.5) {\thicklines \vector(0,-1){5.0}}
\put(22.2,7.5) {\oval(4,4)[l]}
\put(22.2,5.5) {\vector(1,0){0.3}}
\put(25.0,4.7) {\oval(3.6,3.6)[b]}
\put(26.8,4.7) {\vector(0,1){0.3}}
\put(24.0,0.0) {\mbox{(c$^{\prime}$2)}}
%
\put(30.5,5.0) {\line(1,0){9.5}}
\put(30.5,10.0) {\line(1,0){9.5}}
\put(32.5,2.5) {\line(0,1){10}}
\put(37.5,2.5) {\line(0,1){10}}
\put(32.5,2.5) {\thicklines \vector(0,1){5.0}}
\put(37.5,12.5) {\thicklines \vector(0,-1){5.0}}
\put(32.2,7.5) {\oval(4,4)[l]}
\put(32.2,5.5) {\vector(1,0){0.3}}
\put(37.8,7.5) {\oval(4,4)[r]}
\put(37.8,9.5) {\vector(-1,0){0.3}}
\put(34.0,0.0) {\mbox{(c$^{\prime}$3)}}
%
\end{picture}
\begin{picture}(50,20)
\put(0.0,2.5) {\line(1,0){9.5}}
\put(0.0,7.5) {\line(1,0){9.5}}
\put(0.0,12.5) {\line(1,0){9.5}}
\put(2.5,0.0) {\line(0,1){15}}
\put(7.5,0.0) {\line(0,1){15}}
\put(2.5,10.0) {\thicklines \vector(0,1){5.0}}
\put(2.5,5.0) {\thicklines \vector(0,-1){5.0}}
\put(3.0,0.0) {\mbox{(d)}}
%
\put(10.5,2.5) {\line(1,0){9.5}}
\put(10.5,7.5) {\line(1,0){9.5}}
\put(10.5,12.5) {\line(1,0){9.5}}
\put(12.5,0.0) {\line(0,1){15}}
\put(17.5,0.0) {\line(0,1){15}}
\put(12.0,10.0) {\thicklines \vector(0,1){5.0}}
\put(13.0,15.0) {\thicklines \vector(0,-1){5.0}}
\put(11.8,10.0) {\oval(4,4)[l]}
\put(11.8,12.0) {\vector(1,0){0.3}}
\put(11.8,5.0) {\oval(4,4)[l]}
\put(11.8,7.0) {\vector(1,0){0.3}}
\put(13.0,0.0) {\mbox{(d$^{\prime}$1)}}
%
\put(20.5,2.5) {\line(1,0){9.5}}
\put(20.5,7.5) {\line(1,0){9.5}}
\put(20.5,12.5) {\line(1,0){9.5}}
\put(22.5,0.0) {\line(0,1){15}}
\put(27.5,0.0) {\line(0,1){15}}
\put(22.0,5.0) {\thicklines \vector(0,1){5.0}}
\put(23.0,10.0) {\thicklines \vector(0,-1){5.0}}
\put(21.8,10.0) {\oval(4,4)[l]}
\put(21.8,8.0) {\vector(1,0){0.3}}
\put(21.8,5.0) {\oval(4,4)[l]}
\put(21.8,7.0) {\vector(1,0){0.3}}
\put(23.0,0.0) {\mbox{(d$^{\prime}$2)}}
%
\put(30.5,2.5) {\line(1,0){9.5}}
\put(30.5,7.5) {\line(1,0){9.5}}
\put(30.5,12.5) {\line(1,0){9.5}}
\put(32.5,0.0) {\line(0,1){15}}
\put(37.5,0.0) {\line(0,1){15}}
\put(32.0,0.0) {\thicklines \vector(0,1){5.0}}
\put(33.0,5.0) {\thicklines \vector(0,-1){5.0}}
\put(31.8,10.0) {\oval(4,4)[l]}
\put(31.8,8.0) {\vector(1,0){0.3}}
\put(31.8,5.0) {\oval(4,4)[l]}
\put(31.8,3.0) {\vector(1,0){0.3}}
\put(33.0,0.0) {\mbox{(d$^{\prime}$3)}}
\end{picture}
\begin{picture}(20,20)
\put(2.5,2.5) {\circle{2}}
\put(12.5, 2.5) {\circle{2}}
\put(2.5,12.5) {\circle{2}}
\put(12.5,12.5) {\circle{2}}
\put(7.5,7.5) {\circle{2}}
\multiput(0.0,2.5)(1.5,0) {10}{\line(1,0){1}}
\multiput(0.0,7.5)(1.5,0) {10}{\line(1,0){1}}
\multiput(0.0,12.5)(1.5,0) {10}{\line(1,0){1}}
\multiput(2.5,0.0)(0,1.5) {10}{\line(0,1){1}}
\multiput(7.5,0.0)(0,1.5) {10}{\line(0,1){1}}
\multiput(12.5,0.0)(0,1.5){10}{\line(0,1){1}}
\put(0.0,0.0) {\line(1,1){15}}
\put(0.0,5.0) {\line(1,-1){5}}
\put(0.0,10.0) {\line(1,1){5}}
\put(0.0,15.0) {\line(1,-1){15}}
\put(5.0,0.0) {\line(-1,1){5}}
\put(15.0,0.0) {\line(-1,1){15}}
\put(10.0,0.0) {\line(1,1){5}}
\put(10.0,15.0) {\line(1,-1){5}}
\put(6,4){\oval(5,5)[lt]}
\put(2.0,7.5) {\mbox{$J_{\rm eff}$}}
\put(7.0,2.8) {\mbox{$J^{\prime}_{\rm eff}$}}
\put(7.5,2.5) {\oval(7.0,3.0)[t]}
\put(6.0,0.0) {\mbox{(e)}}
\end{picture}
\caption{
Virtual hopping processes up to the fourth order in $t/P$.
Circles indicate charges with no spin specification, while arrows indicate spin.(a)the ground state configuration in the strong coupling limit (zeroth order in $t/P$). (b) the second order perturbation process that gives uniform energy shift $4\dfrac{t^{2}}{3V}$. 
(c) spin pair of the second-nearest neighbor sites and (c$^{\prime}$1) to (c$^{\prime}$3) indicate different processes that contribute in the
the fourth order perturbation. The contribution from each process is $\dfrac{4t^{4}}{(3V)^{2}(4V+U)}$, $\dfrac{4t^{4}}{(3V)^{2}U}$, and $\dfrac{4t^{4}}{(3V)^{2}(4V)}$, respecively.
(d) the spin pair of the third-neighbor sites and (d$^{\prime}$1) to (d$^{\prime}$3) indicate different processes that contribute in the
the fourth order perturbation. The contribution from each process is  $\dfrac{t^{4}}{(3V)^{2}U}$, $\dfrac{4t^{4}}{(3V)^{2}(4V+U)}$, and $\dfrac{t^{4}}{(3V)^{2}U}$, respecively.
(e) The effective $\sqrt{2}\times \sqrt{2}$ magnetic sublattices for the charge-ordered phase. The effective exchange is described as $J_{\rm eff}$ and  $J^{\prime}_{\rm eff}$. }  
\label{fig:exchangefig}
\end{figure}
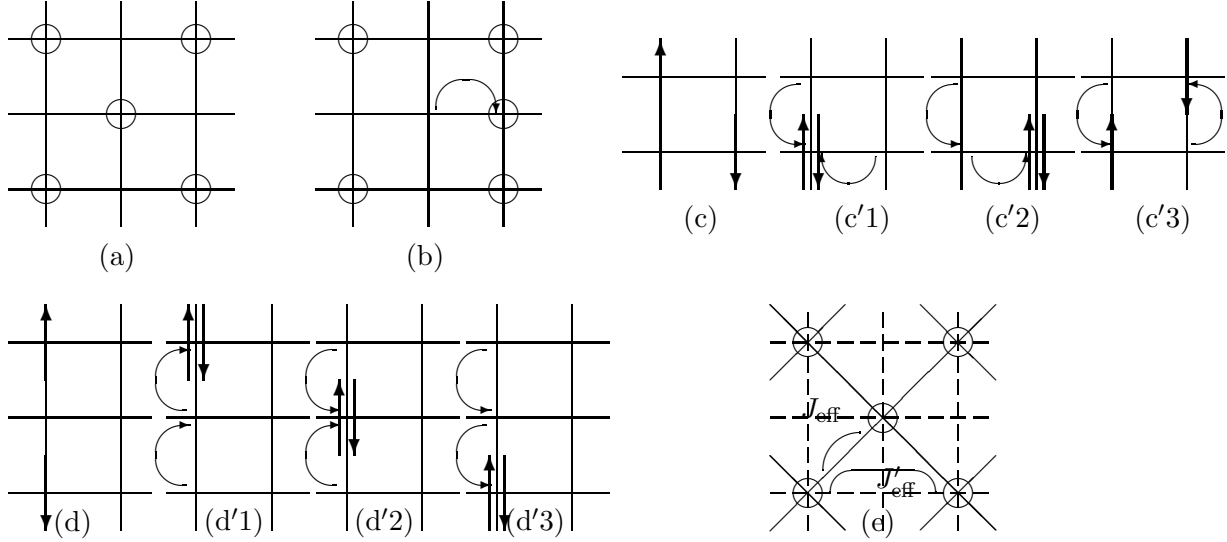

The exchange interaction between localized charges in the checkerboard charge ordered state is derived from virtual processes shown in Fig.~\ref{fig:exchangefig}.
We define the interaction strength parameter \(P \equiv \mbox{min}(U,V)\). We first consider the strong coupling limit where $P/t$ is infinitely large. Then we take account of the effect of finite value of hopping $t$ by order by order in $t/P$. 
In Fig.~\ref{fig:exchangefig},
the second order process (b) merely adds a constant energy shift irrespective of the spin configuration. Non-trivial effect comes from the fourth order perturbation. Contributions from this process differ depending on spin configurations. 
In order to consider effective exchange interaction in the charge ordered phase, we introduce $ \sqrt{2} \times \sqrt{2} $ magnetic lattice. The effective interaction is then expressed in terms of the magnetic lattice as $J_{\rm eff}$ in the nearest neighbor (n.n.) pairs and $J^{\prime}_{\rm eff}$ in the next-nearest neighbor (n.n.n.) or the diagonal directions. Each is obtained as, 
\[ J_{\rm eff}=\dfrac{16t^{4}}{9V^{2}} (\dfrac{1}{U} + \dfrac{1}{4V+U} + \dfrac{1}{4V} ), \]
\[ J^{\prime}_{\rm eff}=\dfrac{4t^{4}}{9V^{2}} (\dfrac{1}{U} + \dfrac{2}{4V+U} ) \]
respectively. Here, the effective Hamiltonian has the form 
\[{\cal H}=J_{\rm eff}\sum_{<i,j>\in \scriptsize \mbox{n.n.}} {\mib S}_{i} \cdot {\mib S}_{j} + J^{\prime}_{\rm eff}\sum_{<i,j>\in \scriptsize \mbox{n.n.n.}} {\mib S}_{i} \cdot {\mib S}_{j}, \]
where $\mib S$ is the spin-$1/2$ operator at the charge-rich site of the charge-ordered phase, and $\sum_{<i,j>\in \scriptsize \mbox{n.n.}}$ indicates the summation over all nearest neighbor pairs, while $\sum_{<i,j>\in \scriptsize \mbox{n.n.n.}}$ indicates that of all next-nearest neighbor pairs (in the sense of $ \sqrt{2} \times \sqrt{2} $ magnetic lattice). 




\end{document}